\documentclass[12pt]{article}

\usepackage[dvipsnames]{xcolor}
\usepackage{amsmath,amsthm,amssymb,natbib,verbatim,cancel, bbm, tikz, tkz-euclide,graphicx,subcaption, float}
\usepackage{caption}
\usetikzlibrary{bending}
\usetikzlibrary{shapes.misc, positioning, arrows,decorations.markings}
 \usetikzlibrary{calc}

 \usepackage{tikz-3dplot}
\usepackage{pgfplots}
\usetikzlibrary{shapes}
\tdplotsetmaincoords{60}{110}

\usepgfplotslibrary{fillbetween}
\usetikzlibrary{intersections}

 \usepackage{hyperref}
\hypersetup{
    colorlinks =true,
    allcolors =NavyBlue,
    allbordercolors =white,
}

\usepackage[margin=1in]{geometry} 
\usepackage{setspace}

\theoremstyle{definition}
\newtheorem{theorem}{Theorem}
\newtheorem*{theorem*}{Theorem}
\newtheorem*{lemma*}{Lemma}
\newtheorem{observation}{Observation}

\newtheorem{corollary}{Corollary}

\newtheorem{lemma}{Lemma}
\newtheorem{example}{Example}
\newtheorem{definition}{Definition}
\definecolor{amber}{rgb}{1.0,0.75,0.0}
\definecolor{aqua}{rgb}{0,1,1}
\definecolor{amaranth}{rgb}{1,.6,.62}

\newcommand\aug{\fboxsep=-\fboxrule\!\!\!\fbox{\strut}\!\!\!}

\title{Identifying Restrictions on the Random Utility Model\footnote{We are grateful to Christopher Chambers and Axel Niemeyer for discussion and helpful comments.}}
\author{Peter P. Caradonna\footnote{Division of the Humanities and Social Sciences, Caltech.  Email: \href{mailto:ppc@caltech.edu}{\nolinkurl{ppc@caltech.edu}}.} \hspace{.1cm} and Christopher Turansick\footnote{Department of Decision Sciences, Bocconi University. Email: \href{mailto:christopher.turansick@unibocconi.it}{\nolinkurl{christopher.turansick@unibocconi.it}}.
}}
\date{\today}

\begin{document}
\onehalfspacing
\maketitle

\begin{abstract}
    We characterize those ex-ante restrictions on the random utility model which lead to identification. We first identify a simple class of perturbations which transfer mass from a suitable pair of preferences to the pair formed by swapping certain compatible lower contour sets.  We show that two distributions over preferences are behaviorally equivalent if and only if they can be obtained from each other by a finite sequence of such transformations. Using this, we obtain specialized characterizations of which restrictions on the support of a random utility model yield identification, as well as of the extreme points of the set of distributions rationalizing a given data set. Finally, when a model depends smoothly on some set of parameters, we show that under mild topological assumptions, identification is characterized by a straightforward, local test.
\end{abstract}

\section{Introduction}

Modern economics is founded on the concept that economic actors take actions to maximize their individual well-being, as described by a preference ranking over outcomes or alternatives.  When such a ranking depends only on known, observable variation in the state of the world, revealed preference theory has shown that these rankings may be recovered from sufficiently rich observational data (e.g.\ \citealt{hurwicz1971integrability, mas1978revealed}; see also \citealt{chambers2021recovering}).  However, when agents' preferences also depend on varying unobservable factors, these preferences, and hence agents' behavior, will appear random.\medskip

In contrast with the positive findings of the classical revealed preference literature in the context of deterministic choice, when preferences are stochastic, it has long been recognized that the distribution over preferences cannot, in general, be uniquely recovered from observed choice frequencies (e.g.\ \citealt{falmagne1978representation, fishburn1998stochastic}).\footnote{In fact, \cite{fishburn1998stochastic} shows that not even the \emph{set} of preferences being randomized over can generally be determined. \cite{turansick2022identification} shows that this set is identified if and only if there is a unique distribution on preferences consistent with the data.} In response, an extensive literature has emerged, that has studied restrictions on the random utility model which allow analysts to uniquely recover the distribution over preferences or parameters of the underlying model (e.g.\ \citealt{luce1959individual, mcfadden1972conditional, gul2006random, apesteguia2017single, yang2023random, suleymanov2024branching}).\footnote{There is also an extensive subliterature in industrial organization which has studied related problems in the context of the invertibility of demand, see e.g. \cite{berry2024nonparametric}.}\medskip

In this paper, we provide a complete characterization of all possible identifying restrictions on the random utility model, for finite consumption environments. We first define a set of simple, local perturbations that transfer mass between particular pairs of related preferences. Crucially, these perturbations transform distributions over preferences in such a manner as to leave all choice probabilities unaffected.  We term such transformations `Ryser swaps,' in light of a connection with the discrete tomography literature (\citealt{ryser1957combinatorial, fishburn1991sets, Kong1999}).\footnote{We discuss this connection in more depth in \autoref{tomographysection}.} We then show that the full class of transforms which preserve all choice probabilities are precisely those generated by the Ryser swaps.\medskip

\begin{example}\label{introex}
    Suppose that an agent has preferences over four pieces of fruit: an apple, a banana, some cherries, or a dragonfruit $\{a,b,c,d\}$, and consider the following four preferences, where each preference is listed in order of descending desirability:\footnote{In other words, $\succ_1$ corresponds to the ordering $a \succ b \succ c \succ d$, and so forth.}
    \begin{table}[h!]
\centering
 \begin{tabular}{c c | c c} 
 
 $\succ_1$ & $\succ_2$ & $\succ_3$ & $\succ_4 $\\ [0.25ex] 
 \hline\hline
  a & b & a & b\\ 
  b & a & b & a\\
  c & d & d & c\\
  d & c & c & d\\ 
 \end{tabular}
\end{table}

\noindent Suppose we have two models of an agent's behavior: the first says that the agent draws their preference from $\mu_{12}$, the uniform distribution supported on $\succ_1$ and $\succ_2$; the second that the preference is drawn uniformly from $\succ_3$ and $\succ_4$, denoted $\mu_{34}$. Note that, for any choice set, the probabilities of this agent choosing a given alternative from this set are identical under either model.\footnote{This observation is originally due to \cite{fishburn1998stochastic}.}\medskip

The key structure is that each of these four preferences can be decomposed into a choice of `initial' ranking (i.e. whether $a \succ b$ or $b \succ a$) and a `terminal' ranking (whether $c \succ d$ or $d \succ c$).  For each preference, both $a$ and $b$ always dominate $c$ and $d$, whatever the respective orderings \emph{within} these groupings are.\footnote{Put differently, these four preferences can be viewed as an element in a product set, where one first fixes a choice of initial ranking, and then a choice of terminal. Two distributions induce the same choice probabilities if and only if both have the same \emph{marginals}, i.e. the probabilities of $a \succ b$ versus $b \succ a$, and probability of $c \succ d$ versus $d\succ c$.} Moreover, $\mu_{34}$ can be obtained from $\mu_{12}$ in a particularly simple fashion: take all the mass on $\succ_1$ (resp. $\succ_2$) and re-assign it to the preference formed by replacing the terminal segment of $\succ_1$ with that of $\succ_2$ (resp. vice-versa).\footnote{Analogously, by reversing this process we could similarly obtain $\mu_{12}$ from $\mu_{34}$.} We term such a transfer of mass between pairs of preferences formed by swapping (compatible) terminal segments a \emph{Ryser swap}; whenever two distributions of preferences are related by a Ryser swap, they must have identical choice probabilities.\medskip

However, many pairs of behaviorally equivalent distributions do not differ from one another by such a swap. For example, the uniform distribution $\mu_{123}$ below is behaviorally equivalent to the uniform $\mu_{456}$, but any Ryser swap between pairs of preferences in this set necessarily either fixes both preferences, or yields at least one preference outside the set.
\begin{table}[h!]
\centering
 \begin{tabular}{c c c | c c c} 
 
 $\succ_1$ & $\succ_2$ & $\succ_3$ & $\succ_4 $ & $\succ_5$ & $\succ_6$\\ [0.25ex] 
 \hline\hline
  a & b & c & a & b & c\\ 
  b & a & d & b & a & d\\
  c & e & b & e & c & b\\
  d & f & a & f & d & a\\ 
  e & c & f & c & f & e\\
  f & d & e & d & e & f
 \end{tabular}
\end{table}

\noindent As such, $\mu_{123}$ and $\mu_{456}$ cannot be related by any transformation of this form.  However, they are related by a \emph{sequence} of Ryser swaps. 

\[
    \begin{array}{ccc}
    \textcolor{blue}{\textrm{a}} & \textcolor{blue}{\textrm{b}} & \textrm{c}\\
    \textcolor{blue}{\textrm{b}} & \textcolor{blue}{\textrm{a}} & \textrm{d}\\ \cline{1-2}
    \textcolor{blue}{\textrm{c}} & \textcolor{blue}{\textrm{e}} & \textrm{b}\\
    \textcolor{blue}{\textrm{d}} & \textcolor{blue}{\textrm{f}} & \textrm{a}\\
    \textcolor{blue}{\textrm{e}} & \textcolor{blue}{\textrm{c}} & \textrm{f}\\
    \textcolor{blue}{\textrm{f}} & \textcolor{blue}{\textrm{d}} & \textrm{e}\\
    \end{array}
    \longrightarrow 
    \begin{array}{ccc}
    \textrm{a} & \textcolor{red}{\textrm{b}} & \textcolor{red}{\textrm{c}}\\
    \textrm{b} & \textcolor{red}{\textrm{a}} & \textcolor{red}{\textrm{d}}\\
    \textrm{e} & \textcolor{red}{\textrm{c}} & \textcolor{red}{\textrm{b}}\\
    \textrm{f} & \textcolor{red}{\textrm{d}} & \textcolor{red}{\textrm{a}}\\\cline{2-3}
    \textrm{c} & \textcolor{red}{\textrm{e}} & \textcolor{red}{\textrm{f}}\\
    \textrm{d} & \textcolor{red}{\textrm{f}} & \textcolor{red}{\textrm{e}}\\
    \end{array}
    \longrightarrow
    \begin{array}{ccc}
    \textrm{a} & \textrm{b} & \textrm{c}\\
    \textrm{b} & \textrm{a} & \textrm{d}\\
    \textrm{e} & \textrm{c} & \textrm{b}\\
    \textrm{f} & \textrm{d} & \textrm{a}\\
    \textrm{c} & \textrm{f} & \textrm{e}\\
    \textrm{d} & \textrm{e} & \textrm{f}\\
    \end{array}
\]

\noindent Starting from $\mu_{123}$, first transfer all the mass from $\succ_1$ and $\succ_2$ to the pair of preferences formed by swapping their terminal segments $cdef$ and $efcd$. Label these resulting preferences $\succ_1'$ and $\succ_2'$.  Then repeat this operation, transferring all mass from $\succ_2'$ and $\succ_3$ to the pair obtained by swapping their terminal segments, $ef$ and $fe$ respectively.  This yields $\mu_{456}$, as desired. \hfill $\blacksquare$
\end{example}

More generally, we show that any two distributions are behaviorally indistinguishable if and only if they can be obtained from one another by a finite sequence of (weighted) swaps of this form.\footnote{`Weighted' simply meaning that these swaps need not transfer \emph{all} the mass on the initial pair of preferences to the swapped pair.}  We show that every sequence of weighted Ryser swaps can be identified with particular signed measure over preferences, and the set of all such measures forms a linear subspace $\mathcal{R}$, which we term the `Ryser subspace.'  Using this observation, we provide a geometric characterization of identifying restrictions: an arbitrary subset of the set of distributions over preferences is identified if and only if its intersection with each translate of the Ryser subspace is at most singleton.\medskip

We then consider how our results specialize to more structured, common classes of restrictions.  Perhaps the most natural candidate are restrictions on the set of preferences a distribution can randomize over.  We prove that such support restrictions are identifying if and only if every non-zero vector in the Ryser subspace places weight on some preference outside the allowed support.\medskip

We leverage this to provide a characterization of the extremal rationalizations of consistent choice data, under arbitrary support restrictions.  Given a (generally non-identifying) support restriction, the set of allowable distributions over preferences which rationalize a particular set of observations forms a polytope. We show that a rationalizing distribution is an extreme point of this set if and only if \emph{its} support has the property that every vector in the Ryser subspace places non-zero weight outside it.\medskip

Finally, we turn to parametric forms of restriction.  Here, we consider sets of distributions varying smoothly with some finite-dimensional vector of parameters.  Our earlier results provide an immediate global characterization: a parametric model is identified if and only if no two vectors of parameters induce distributions which are related by a finite sequence of weighted Ryser swaps. However, we show that under mild, topological restrictions on the set of random choice rules consistent with the model, global identification is fully characterized by a purely \emph{local}, full-rank condition on a particular Jacobian matrix.  We show that geometrically, this is equivalent to requiring that no translate of the Ryser subspace is ever tangent to the surface of distributions traced out by the parameterization function. As an application, we show that any mixture model is identified if and only if it is so locally.\medskip

The rest of this paper proceeds as follows. In \autoref{sec:Model}, we formally introduce the random utility model and other preliminaries. In \autoref{sec:geometry}, we provide a formal construction of the Ryser subspace, present our main result, and provide a discussion. In \autoref{sec:fullSub}, we characterize identifying support restrictions, and in \autoref{sec:Parm}, we consider the case of parametric random utility models. We briefly conclude in \autoref{sec:conc}.

\section{Related Literature}
The two papers most closely related to ours are \citet{fishburn1998stochastic}  and \citet{turansick2022identification}. The first explicit example of an unidentified random utility model appeared in \citet{fishburn1998stochastic}; this construction turns out to form the basis of our definition of conjugate square.  \citet{turansick2022identification} characterizes which distributions are the unique random utility models with their choice probabilities; see also \cite{doignon2023adjacencies}.  As discussed in  \autoref{Ex:3by3}, the results of \citet{turansick2022identification} follow as corollaries of our main results here.\medskip

 \citet{doignon2023adjacencies} study which pairs of rational choice functions form adjacent vertices on the random utility polytope. They find that adjacency of two choice functions corresponds to the two preferences they represent having either a common, non-trivial initial segment, and different terminal segments, or a common non-trivial terminal segment with differing initial segments.  Using this characterization, Doignon and Saito are able to recover the results of \citet{turansick2022identification} purely in terms of this adjacency relation. We discuss the  formal connection between our results and the structure of the random utility polytope in \autoref{App:Polytope}.\medskip
 
 A number of papers consider restrictions on the random utility model for purposes of obtaining identification. \citet{apesteguia2017single} considers random utility models whose support obeys the single-crossing property with respect to some exogenous ordering on alternatives $\trianglerighteq$, and show that any data set is consistent with at most one such model.  Extensions and variations on this idea are considered in \citet{yildiz2022foundations} and \citet{filiz2023progressive}. \cite{manzini2018dual} restrict to random utilities which only depend on two underlying states of the world. Manzini and Mariotti show that, in this case, the underlying distribution over preferences can be generically recovered. \citet{suleymanov2024branching} studies the branching-independent random utility model. The model recovers identification by restricting to statistically independent maps from contour sets to preferences. \citet{honda2021random} considers a random cravings model where an agent is endowed with a base preference and cravings which arrive randomly, and obtains identification under certain monotonicity restrictions.\medskip
 
 In practice, it is common to specify random utility model parametrically.  Perhaps the most well-known identified parametric restriction is due \citet{luce1959individual}.  This model is behaviorally equivalent to the logit model of \citet{mcfadden1972conditional}.\footnote{See also \cite{sandomirskiy2023decomposable} for a recent characterization.} More recently, \citet{chambers2024weighted} considers an extension of the Luce model including both salience and utility in its parameterization which preserves identification.\medskip
 
While outside the scope of this paper, a related strand of literature instead considers the identification problem under various assumptions on data and observability. \cite{dardanoni2020inferring} and \citet{dardanoni2023mixture} consider mixture choice data which allows analysts to connect the choices of a single agent across menus. In contrast, \cite{azrieli2022marginal} consider a weaker type of data where analysts observe the frequency of choosing an alternative and the frequency a menu is realized but do not observe the frequency with which alternative is chosen from a menu. They show that identification in this setting fares worse than with standard stochastic choice data.\medskip

A number of papers also consider identification of random utility models on infinite domains. \citet{gul2006random} shows that the random expected utility model is identified.  Similarly, the random quasilinear utility model is identified  \citep{williams1977formation, daly1979identifying, yang2023random}.  Our results are not directly applicable to these models due to their requirement of an infinite domain.

\section{Model and Primitives}\label{sec:Model}

Let $X$ denote a fixed, finite set of alternatives over which an individual has a preference.  A {\bf preference} is a linear order on $X$; we denote the set of all preferences by $\mathcal{L}$.\footnote{A linear order is a complete, transitive, and antisymmetric binary relation on $X$.} To conserve on notation, we will often denote preferences in list form, i.e.\ `$abcd$' to denote the preference $a \succ b \succ c \succ d$. For any preference $\succ$ and any $1 \le k \le \vert X \vert$, we refer to the $k$-{\bf initial} and $k$-{\bf terminal} segments of $\succ$ as the ordered strings consisting of the $k$ most-preferred alternatives under $\succ$ and the $\vert X \vert - k$ least-preferred alternatives, and we denote these by $s^\uparrow_k(\succ)$ and $s^\downarrow_k(\succ)$ respectively. For example, if $\succ$ denotes the order $abcde$, then:
\[
    s^\uparrow_2(\succ) = ab \quad \textrm{ and } \quad s^\downarrow_2(\succ) = cde.
\]

A function $\rho: X \times 2^X \setminus \{\varnothing\} \to [0,1]$ defines a {\bf random choice rule} if, for all non-empty subsets $A \subseteq X$,
\[
    \sum_{x \in A} \rho(x,A) = 1.
\]
Random choice rules are assumed observable; they constitute the basic data in our identification problem.  For any finite set $A$, we use $\Delta(A)$ to denote the set of probability measures over $A$. A {\bf random utility model} (or simply a model) is an arbitrary subset $\mathcal{M} \subseteq \Delta(\mathcal{L})$. We say a random choice rule is rationalizable by a model $\mathcal{M}$ (or $\mathcal{M}$-rationalizable) if there exists some probability distribution $\mu \in \mathcal{M}$ such that, for all $x \in A \subseteq X$, we have:
\[
    \rho(x,A) = \mu\{\succ \, \in \mathcal{L}\;  \vert \; x \textrm{ is maximal in } A \} = \sum_{\succ \in \mathcal{L}} \mu(\succ) \mathbbm{1}_{\{x \; \succ \,A \setminus x\}}.
\]
Under the interpretation of a model $\mathcal{M}$ as describing  a set of possible makeups of a heterogeneous society, a collection of observed choice frequencies $\rho$ are $\mathcal{M}$-rationalizable if and only if they could arise as the distribution of constrained-optimal outcomes according to some composition $\mu \in \mathcal{M}$ of the population.\medskip

Given a model $\mathcal{M}$, we say that two measures $\mu, \nu \in \mathcal{M}$ are {\bf observationally equivalent} if, for all $x \in A \subseteq X$,
\begin{equation}\label{behavioralequiv}
    \mu\{\succ \, \in \mathcal{L}\;  \vert \; x \textrm{ is maximal in } A \} = \nu\{\succ \, \in \mathcal{L}\;  \vert \; x \textrm{ is maximal in } A \},
\end{equation}
i.e. they generate identical choice frequencies on every choice set $\varnothing \subsetneq A \subseteq X$.  Finally, we say a model $\mathcal{M}$ is  {\bf identified} if it contains no pair of distinct, observationally equivalent measures.

\section{The Geometry of Identification}\label{sec:geometry}

Consider first the unrestricted model $\mathcal{M} = \Delta(\mathcal{L})$.  It has been recognized since \citet{barbera1986falmagne} and \citet{fishburn1998stochastic} that the unrestricted random utility model fails to be identified.  Thus, to each distribution $\mu \in \Delta(\mathcal{L})$, there is some (possibly singleton) equivalence class $[\mu]$ of behaviorally indistinguishable distributions. \medskip

Let $\mathcal{P}$ denote the set of random choice rules on $X$, and define the mapping $\Phi: \Delta(\mathcal{L}) \to \mathcal{P}$ via:
\begin{equation}\label{measuretochoiceprobmap}
    \Phi(\mu)_{(x,A)} = \begin{cases}
         \mu\{\succ \, \in \mathcal{L}\;  \vert \; x \textrm{ is maximal in } A \} & \textrm{ if } x \in A\\
         0 & \textrm{ if } x \not \in A.
    \end{cases}
\end{equation}
By equation \eqref{behavioralequiv}, two distributions $\mu,\nu \in \Delta(\mathcal{L})$ are behaviorally indistinguishable if and only if $\Phi(\mu) = \Phi(\nu)$.  However, as $\Phi$ is linear, it follows that geometrically each equivalence class of behaviorally indistinguishable distributions $[\mu]$ is formed by intersecting a translate of some fixed, linear subspace of $\mathbb{R}^\mathcal{L}$ with the simplex $\Delta(\mathcal{L})$.\footnote{Where $\mathbb{R}^\mathcal{L}$ denotes the space of all signed measures over linear orders on $X$.}  The primary objective of this section will be to provide a characterization of this subspace.

\subsection{Separable Pairs and the Ryser Subspace}

A basic criterion for behavioral indistinguishability was introduced by \cite{falmagne1978representation}, in his work on rationalizable random choice rules. Formally, for any $x \in A \subseteq X$, let:
\[
    U(x,A) = \{\succ \in \mathcal{L} \; \vert \; A \setminus x \textrm{ is the strict upper contour set at $x$}\}.
\]
Falmagne established that two distributions in $\Delta(\mathcal{L})$ are behaviorally indistinguishable if and only if they place equal measure on each set $U(x,A)$.

\begin{theorem*}[\citealt{falmagne1978representation}]
    Let $\mu, \nu \in \Delta(\mathcal{L})$. Then $\mu$ and $\nu$ are behaviorally equivalent if and only if, for all $x \in A \subseteq X$,
    \[
        \mu \big[U(x,A)\big] = \nu\big[U(x,A)\big].
    \]
\end{theorem*}
By Falmagne's result, two distributions are behaviorally indistinguishable if and only if they induce equivalent distributions over upper contour sets.  Thus, given some measure $\mu \in \Delta(\mathcal{L})$, to obtain a behaviorally equivalent distribution $\nu \in \Delta(\mathcal{L})$, we seek to take mass from some collection of orders in the support of $\mu$ and reassign it to a different set of orders that, in some sense, `shuffle' the upper contour sets of this collection.\medskip

To formalize this idea, we say a pair of preferences $\succ,\succ'\, \in \mathcal{L}$ form a {\bf separable pair} if, for some $2 \le k \le \vert X \vert - 2$:
\begin{itemize}
    \item[(i)] The preferences agree on the set of $k$-most preferred alternatives, but not on their rankings (i.e.\ $s^\uparrow_k(\succ) \neq s^\uparrow_k(\succ')$); and
    \item[(ii)] The preferences agree on the set of $(\vert X \vert - k)$-least preferred alternatives, but not on their rankings (i.e.\ $s^\downarrow_k(\succ) \neq s^\downarrow_k(\succ')$).
\end{itemize}
A separable pair of preferences agree on the \emph{set} of $k$-most (and $(\vert X \vert - k)$-least) desirable alternatives, but, crucially, not on the orderings within these sets.  This means that by swapping the $k$-initial segments of these two preferences, we obtain two new, distinct preferences which possess \emph{the same upper contour sets}.\medskip


\begin{example}[\citealt{fishburn1998stochastic}]\label{fishex}
Suppose $X = \{a,b,c,d\}$, and consider the following four linear orders:\bigbreak
\begin{table}[h!]
\centering
 \begin{tabular}{c c c c} 
 
 $\succ_1$ & $\succ_2$ & $\succ_3$ & $\succ_4 $\\ [0.25ex] 
 \hline\hline
  a & b & a & b\\ 
  b & a & b & a\\
  c & d & d & c\\
  d & c & c & d\\ 
 \end{tabular}
\end{table}

\noindent Note that $\succ_1$ and $\succ_2$ (as well as $\succ_3$ and $\succ_4$) form a separable pair, with $k=2$. In fact, we can obtain $\succ_3$ and $\succ_4$ from $\succ_1$ and $\succ_2$ by simply swapping initial segments:
\[
    \begin{aligned}
        \succ_3 & = s^\uparrow_2(\succ_1)\cdot s^\downarrow_2(\succ_2)\\
        \succ_4 & = s^\uparrow_2(\succ_2)\cdot s^\downarrow_2(\succ_1),
    \end{aligned}
\]
where $s^\uparrow_k(\succ_i) \cdot s^\downarrow_k(\succ_j)$ denotes the preference formed by concatenation of these segments.  Thus by Falmagne's theorem, it follows that the uniform distribtion on $\{\succ_1, \succ_2\}$ and the uniform distribution on $\{\succ_3,\succ_4\}$ are behaviorally identical.\footnote{In fact, the uniform distribution on $\{\succ_1,\ldots, \succ_4\}$ is also behaviorally indistinguishable from either of these.} \hfill $\blacksquare$
\end{example}

The behavioral indeterminacy arising from separable pairs may equivalently be regarded as a consequence of the fact that a joint distribution is not, in general, uniquely determined by its marginals.  Suppose that $\succ_1, \succ_2$ form a separable pair for some $2 \le k \le \vert X \vert - 2$, and let $\succ_3$ and $\succ_4$ denote the (distinct) pair of preferences obtained by swapping the $k$-initial segments of $\succ_1$ and $\succ_2$. We say that the preferences $\{\succ_3, \succ_4\}$ obtained in this manner are {\bf conjugate} to the pair $\{\succ_1,\succ_2\}$; we term a pair of such pairs, $\{\succ_1,\ldots \succ_4\}$, a {\bf conjugate square}. Any conjugate square can be identified with the product $\{s^\uparrow_k(\succ_1), s^\uparrow_k(\succ_2)  \} \times \big\{s^\downarrow_k(\succ_1), s^\downarrow_k(\succ_2) \big\}$ via the mapping:
\[
    \big(s^\uparrow_k(\succ_i), s^\downarrow_k(\succ_j)\big) \mapsto s^\uparrow_k(\succ_i) \cdot s^\downarrow_k(\succ_j).
\]
For any such tuple, the following simple corollary of Falmagne's theorem implies that two distributions supported on a conjugate pair $\{\succ_1,\ldots, \succ_4\}$ are behaviorally identical if and only if their marginals coincide.

\begin{figure}
\centering
\begin{subfigure}[t]{.45\textwidth}
\begin{tikzpicture}[scale = .65]
        \draw[thick] (0,0)--(0,6)--(6,6)--(6,0)--cycle;
        \draw[thick] (0,0)--(6,0)--(6,3)--(0,3)--cycle;
        \draw[thick] (0,0)--(0,6)--(3,6)--(3,0)--cycle;

        \node[left] at (-.5,1.5) { \scriptsize $\big\{s^\uparrow_2(\succ) = ba \big\}$};
        \node[left] at (-.5,4.5) {\scriptsize  $\big\{s^\uparrow_2(\succ) = ab \big\}$};d

        \node[above] at (1.5, 6.5) {\scriptsize  $\big\{s^\downarrow_2(\succ) = cd \big\}$};
        \node[above] at (4.5, 6.5) { \scriptsize $\big\{s^\downarrow_2(\succ) = dc \big\}$};

        \node at (1.5,4.5) { $\succ_1$};
        \node at (1.5,1.5) { $\succ_4$};
        \node at (4.5,4.5) { $\succ_3$};
        \node at (4.5,1.5) { $\succ_2$};

    \end{tikzpicture}
        \subcaption{The conjugate square $\{\succ_1,\ldots, \succ_4\}$ from \autoref{fishex}, viewed as a product space. Joint measures are behaviorally indistinguishable if and only if their marginals coincide.}
\end{subfigure}
\begin{subfigure}[t]{.45\textwidth}
    \centering
    \includegraphics[width=0.75\linewidth]{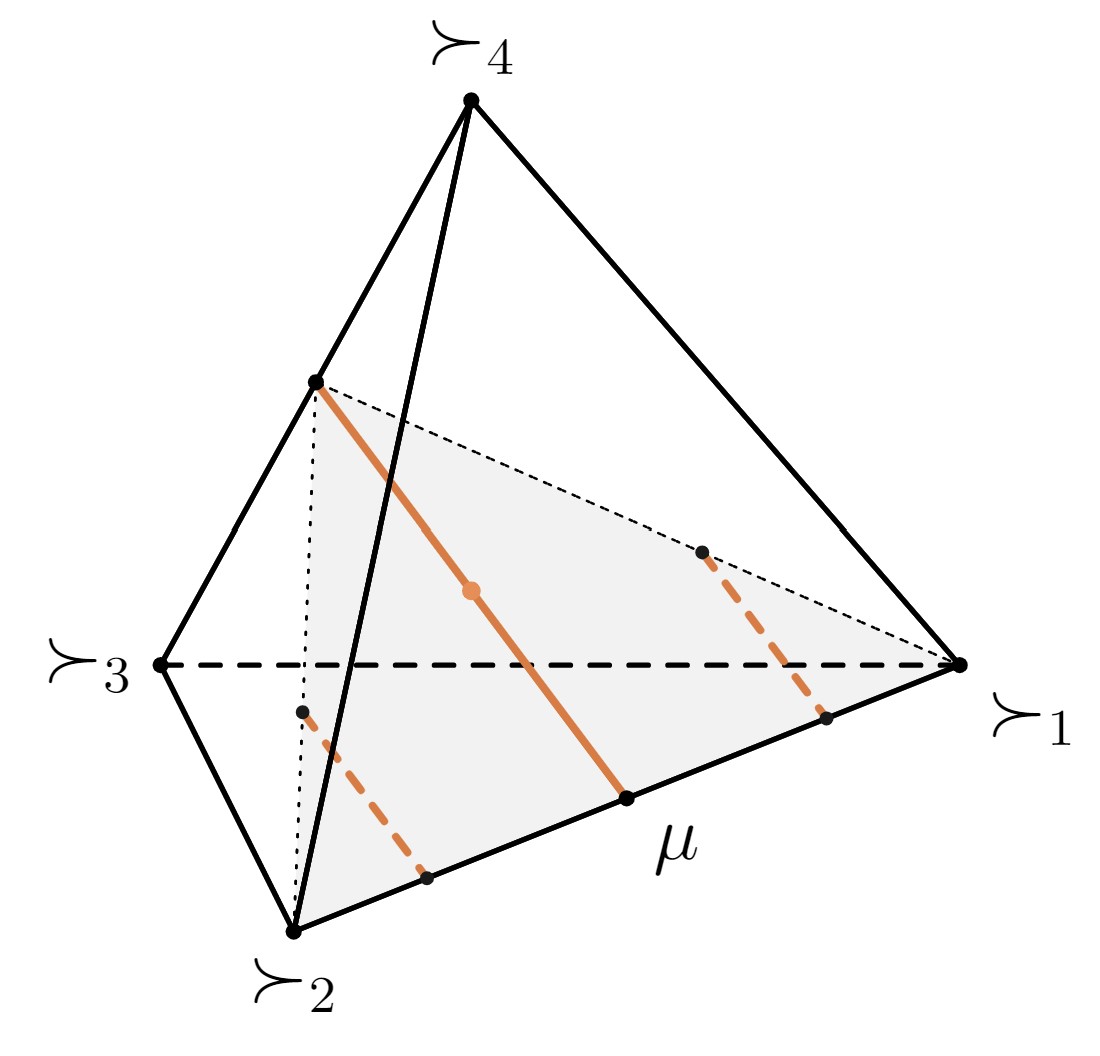}
\subcaption{The set of distributions behaviorally equivalent to $\mu$ (orange, solid), as well as two other behavioral equivalence classes (orange, dotted).}
\end{subfigure}
\caption{The behavioral equivalence classes are given by the intersection of translates of the Ryser subspace with the simplex $\Delta\big(\{\succ_1,\ldots, \succ_4\}\big)$. The uniform distribution $\mu$ on $\{\succ_1,\succ_2\}$ is behaviorally equivalent to the the uniform distribution on its conjugate, $\{\succ_3,\succ_4\}$ (and to the uniform over all four preferences). As $\mu$ varies along the edge connecting $\succ_1$ and $\succ_2$, the behavioral equivalence classes trace out the shaded gray region.}
\label{fig:ProdAndSimp}
\end{figure}
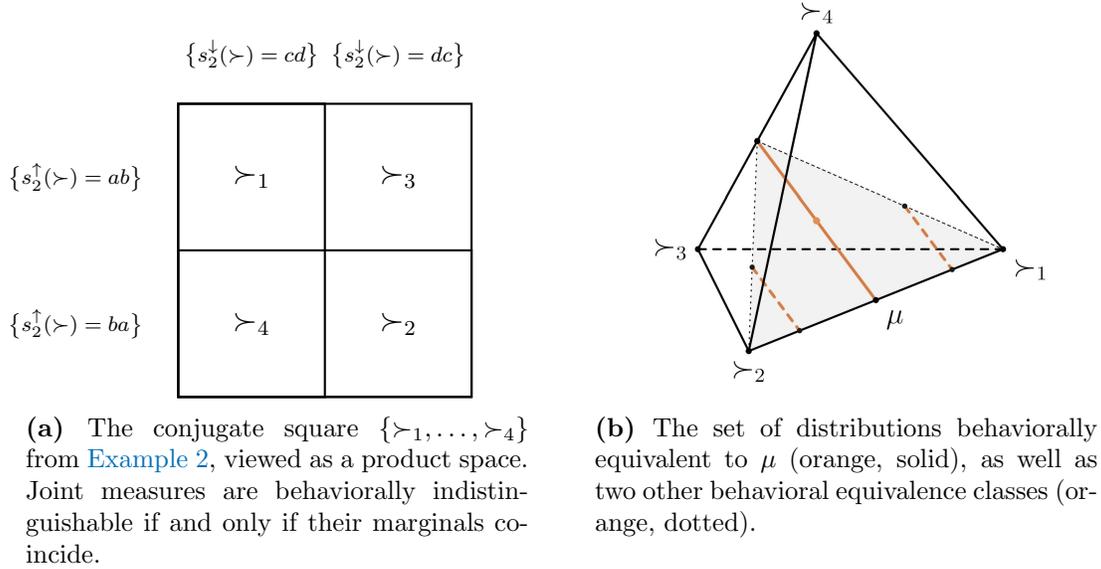

\begin{corollary}\label{marginalequiv}
    Let $\{\succ_1,\ldots, \succ_4\}$ be a conjugate pair, for some $2 \le k \le \vert X \vert -2$. Then for $\mu, \nu \in \Delta\big( \{\succ_1,\ldots, \succ_4\}\big)$ the following are equivalent:
    \begin{itemize}
        \item[(i)] $\mu$ and $\nu$ are behaviorally indistinguishable.
        \item[(ii)] For every $x \in A \subseteq X$, $\mu\big[U(x,A)\big] = \nu\big[U(x,A)\big]$.
        \item[(iii)] The marginals of $\mu$ and $\nu$ over $\{s^\uparrow_k(\succ_1), s^\uparrow_k(\succ_2)  \}$ and $\{s^\downarrow_k(\succ_1), s^\downarrow_k(\succ_2) \}$ coincide, i.e.:
    \[  
        \mu \big\{\succ \vert \, s^\uparrow_k(\succ) =       s^\uparrow_k(\succ_i)\big\} = \nu \big\{\succ \vert \,    s^\uparrow_k(\succ) = s^\uparrow_k(\succ_i)\big\}
    \]
    and
    \[
        \mu \big\{\succ \vert \, s^\downarrow_k(\succ) =       s^\downarrow_k(\succ_j)\big\} = \nu \big\{\succ \vert \,    s^\downarrow_k(\succ) = s^\downarrow_k(\succ_j)\big\},
    \]
    for $i \in \{1,2\}$ and $j \in \{3,4\}$.
    \end{itemize}
\end{corollary}

\noindent We say a signed measure $R \in \mathbb{R}^\mathcal{L}$ defines a {\bf Ryser swap} if:
\[
    R = \mathbbm{1}_{\{\succ_3,\succ_4\}} - \mathbbm{1}_{\{\succ_1,\succ_2\}}
\]
for some conjugate square $\{\succ_1, \ldots, \succ_4 \}$. Similarly, a signed measure defines a {\bf weighted Ryser swap} if it is proportional to a Ryser swap.  A weighted Ryser swap (with weight $\alpha$) may be viewed as a perturbation taking $\alpha$ mass away from $\{\succ_1,\succ_2\}$ and reassigning it to the conjugate pair $\{\succ_3,\succ_4\}$. By \autoref{marginalequiv}, this may equivalently be interpreted as separating out from an initial measure $\mu \in \Delta(\mathcal{L})$  a `joint' distribution on a product set, applying a marginal-preserving transformation to it, then recombining this modified joint distribution with the remaining mass in $\mu$ to obtain a modified distribution $\mu'$.\medskip

We define the {\bf Ryser subspace} $\mathcal{R}\subseteq \mathbb{R}^\mathcal{L}$ as the span of the Ryser swaps:
\[
    \mathcal{R} = \textrm{span}\big\{R \in \mathbb{R}^\mathcal{L} \; \vert \; R \textrm{ is a Ryser swap}\big\}.
\]
Every signed measure in $\mathcal{R}$ may be regarded as a composite transformation which applies a sequence of mass-transferring perturbations to a given distribution.  Formally, these measures consist precisely of the transformations generated by the (unweighted) Ryser swaps. It follows from Falmagne's theorem that for any $R \in \mathcal{R}$ and any $\mu \in \Delta(\mathcal{L})$, so long as $\mu + R \in \Delta(\mathcal{L})$ too, then $\mu$ and $\mu+R$ are behaviorally equivalent probability distributions. Our next theorem shows that in fact, \emph{every} pair of behaviorally equivalent distributions over preferences are related in this manner.

\begin{theorem}\label{ryserspacetheorem}
    Let $\mathcal{M} \subseteq \Delta(\mathcal{L})$ be arbitrary.  Then the following are equivalent:
    \begin{itemize}
        \item[(i)] $\mathcal{M}$ is identified.
        \item[(ii)] For all $\mu \in \mathcal{M}$,
        \[
            \big(\mu + \mathcal{R}\big) \cap \mathcal{M} = \{\mu\}.
        \]
    \end{itemize}
\end{theorem}

\autoref{ryserspacetheorem} says that not only do the Ryser swaps preserve choice probabilities, but in fact they generate the set of \emph{all} transformations which do so. As an easy consequence of this result, we also obtain a characterization of the sets of distributions belonging to a model which generate equivalent sets of choice probabilities.

\begin{corollary}\label{identifiedsetcorr}
    For all $\mu \in \mathcal{M}$:
    \[
        \big\{\nu \in \mathcal{M} \; \vert \; \nu \textrm{ is behaviorally equivalent to } \mu\} = \big(\mu + \mathcal{R}\big) \cap \mathcal{M}.
    \]
\end{corollary}

This provides a complete characterization of which distributional restrictions on the random utility model are identifying.  As our next example shows, it also allows us to straightforwardly recover existing results from the literature as direct corollaries.

\begin{example}\label{Ex:3by3}
    Consider the unrestricted model $\mathcal{M} = \Delta(\mathcal{L})$.  In the language of this paper, \cite{turansick2022identification} proved that $\mu \in \Delta(\mathcal{L})$ is uniquely determined by its choice probabilities if and only if the support of $\mu$ contains no separable pair.\footnote{Alternatively, this may equivalently be obtained as an immediate consequence of \autoref{identifiedsetcorr}.}\medskip

    In light of, e.g., \autoref{marginalequiv}, this is clearly necessary. However, the proof in \cite{turansick2022identification} relied crucially on the fact that $\mathcal{M}$ was unrestricted, and hence one could always find distributions in $\mathcal{M}$ that differ from some candidate $\mu$ only by shifting some mass from a separable pair to its conjugate.  In contrast, \autoref{ryserspacetheorem} constitutes a significant generalization, characterizing identification for arbitrary models, no matter how complex their defining system of restrictions may be. For example, suppose $X = \{a,b,c,d,e,f\}$, and recall the preferences from \autoref{introex}:
    \begin{table}[h!]
\centering
 \begin{tabular}{c c c c c c} 
 
 $\succ_1$ & $\succ_2$ & $\succ_3$ & $\succ_4 $ & $\succ_5$ & $\succ_6$\\ [0.25ex] 
 \hline\hline
  a & b & c & a & b & c\\ 
  b & a & d & b & a & d\\
  c & e & b & e & c & b\\
  d & f & a & f & d & a\\ 
  e & c & f & c & f & e\\
  f & d & e & d & e & f
 \end{tabular}
\end{table}

\noindent Let $\mathcal{M} = \Delta\big(\{\succ_1,\ldots, \succ_6\}\big)$, and consider the uniform distribution $\mu$.  It may straightforwardly be verified that $\mu$ is behaviorally equivalent to both the uniform distribution $\mu_{123}$ on $\{\succ_1,\succ_2, \succ_3\}$ and the uniform distribution $\mu_{456}$ on $\{\succ_4,\succ_5, \succ_6\}$. However, $\mathcal{M}$ does not contain any conjugate square in its support: applying any single weighted Ryser swap, to any distribution in $\mathcal{M}$, necessarily results in a distribution not belonging to $\mathcal{M}$. As a consequence, the arguments of \cite{turansick2022identification} are not applicable. Nonetheless, \autoref{ryserspacetheorem} guarantees that all three of these distributions can be obtained from one another by an appropriate \emph{sequence} of such transformations.\hfill $\blacksquare$
\end{example}

\subsection{Proof Sketch}
In light of \autoref{ryserspacetheorem}, two distributions over preferences are observationally equivalent if and only if they may be obtained from one another by applying a finite sequence of weighted Ryser swaps. However, \autoref{ryserspacetheorem} is silent on how to construct such sequences.\medskip

Given two observationally equivalent distributions, the crux of our proof is to establish the purely combinatorial proposition that any preference in the support of either distribution can be constructed by iteratively swapping the terminal segments of preferences in the support of the other. We term this procedure `zippering.'  To illustrate, recall the two distributions considered \autoref{Ex:3by3}. The first, $\mu_{123}$, corresponded to the uniform distribution on $\succ_1$ to $\succ_3$, while the latter, $\mu_{456}$ to the uniform distribution on $\succ_4$ to $\succ_6$.  We seek to construct the preference $\succ_1$ out of a sequence of terminal segment swaps, starting with those preferences in the support of $\mu_{456}$.\medskip

Our zippering procedure starts by finding a preference $\succ_i$ in the support of $\mu_{456}$ such that $s_k^\uparrow(\succ_1) = s_k^\uparrow(\succ_i)$ for some $2 \le k \le 4$. Falmagne's theorem ensures such a $k$ must exist. In this example, we see that $s_2^\uparrow(\succ_1) = s_2^\uparrow(\succ_4)$.

    \begin{table}[h!]
\centering
 \begin{tabular}{c c c | c c c} 
 
 $\succ_1$ & $\succ_2$ & $\succ_3$ & $\succ_4 $ & $\succ_5$ & $\succ_6$\\ [0.25ex] 
 \hline\hline
  {\color{blue} a} &  b & c & {\color{blue} a} & b & c\\ 
  {\color{blue} b} &  a & d & {\color{blue} b} & a & d\\
  {\color{blue} c} &  e & b & {\color{red} e} & c & b\\
  {\color{blue} d} &  f & a & {\color{red} f} & d & a\\ 
  {\color{blue} e} &  c & f & {\color{red} c} & f & e\\
  {\color{blue} f} &  d & e & {\color{red} d} & e & f
 \end{tabular}
\end{table}

\noindent Note that $\{a,b\}$ is the upper contour set at $c$ under $\succ_1$, though not $\succ_4$. By \cite{falmagne1978representation} once again, we are able to ensure there is some preference in the support of $\mu_{456}$ with this upper contour set at $c$ as well, here $\succ_5$.  The two preferences in the support of $\mu_{456}$ obtained this way, $\succ_4$ and $\succ_5$, must form a separable pair.  Our zippering procedure then replaces this pair with their conjugate, resulting in a modified set of preferences, $\{\succ_4', \succ_5', \succ_6\}$.

    \begin{table}[h!]
\centering
 \begin{tabular}{c c c | c c c} 
 
 $\succ_1$ & $\succ_2$ & $\succ_3$ & $\succ_4' $ & $\succ_5'$ & $\succ_6$\\ [0.25ex] 
 \hline\hline
  {\color{blue} a} &  b & c & {\color{blue} a} & b & c\\ 
  {\color{blue} b} &  a & d & {\color{blue} b} & a & d\\
  {\color{blue} c} &  e & b & {\color{blue} c} & e & b\\
  {\color{blue} d} &  f & a & {\color{blue} d} & f & a\\ 
  {\color{blue} e} &  c & f & {\color{red} f} & c & e\\
  {\color{blue} f} &  d & e & {\color{red} e} & d & f
 \end{tabular}
\end{table}

Within this new set, $\succ_4'$ now has the property that $s_{k'}^\uparrow(\succ_1) = s_{k'}^\uparrow(\succ_4')$ for $k' = 4 > 2$.\footnote{And indeed this is the largest $k'$ such that this remains true, as $s_5^\uparrow(\succ_1) \neq s_5^\uparrow(\succ_4')$.}  Thus, repeating the logic from above, we obtain that $\succ_4'$ and $\succ_6$ are separable, and hence their terminal segments may be swapped. Doing so we obtain $\{\succ_4'', \succ_5', \succ_6'\}$, where now $\succ_4'' \, = \,\succ_1$ as desired.\medskip

    \begin{table}[h]
\centering
 \begin{tabular}{c c c | c c c} 
 
 $\succ_1$ & $\succ_2$ & $\succ_3$ & $\succ_4'' $ & $\succ_5'$ & $\succ_6'$\\ [0.25ex] 
 \hline\hline
  {\color{blue} a} &  b & c & {\color{blue} a} & b & c\\ 
  {\color{blue} b} &  a & d & {\color{blue} b} & a & d\\
  {\color{blue} c} &  e & b & {\color{blue} c} & e & b\\
  {\color{blue} d} &  f & a & {\color{blue} d} & f & a\\ 
  {\color{blue} e} &  c & f & {\color{blue} e} & c & f\\
  {\color{blue} f} &  d & e & {\color{blue} f} & d & e
 \end{tabular}
\end{table}

We refer to this procedure as `zippering' as at each stage, we start with some preference which agrees with $\succ_1$ on an initial segment, then by swapping terminal segments, obtain a preference which agrees with $\succ_1$ on a strictly longer initial segment. By `zippering' bits of our other preferences to this approximation of $\succ_1$, we iteratively improve it until we converge to to an exact copy. While we have not emphasized it here, the other crucial aspect of this procedure is it guarantees that we may choose weighted Ryser swaps, corresponding to these reassignments of terminal segments, in such a manner as to guarantee that we shift precisely the right amount of mass onto our duplicate of $\succ_1$.  Together, these properties allow us to guarantee we can always fully recover one distribution from the other.

\subsection{Relation to Discrete Tomography}\label{tomographysection}

Discrete tomography considers the problem of reconstructing a geometric object from various collections of lower-dimensional projections.\footnote{For a textbook treatment, see e.g.\ \cite{herman2012discrete}.}  One problem that has garnered considerable attention is that of characterizing the sets of $0$-$1$ matrices with given row and column sums (e.g.\ \citealt{fishburn1991sets}). In probabilistic language, this may be regarded as the problem of characterizing which compactly-supported uniform distributions on $\mathbb{Z} \times \mathbb{Z}$ have equal marginals.\medskip

\cite{ryser1957combinatorial} provided a complete solution to this problem in terms of a switching operation.  Given any such matrix $M$, Ryser considered `rectangular' configurations of entries $ M_{ij}, M_{ij'}, M_{i'j}, M_{i'j'}$, where either precisely $M_{ij}$ and $M_{i'j'}$ are equal to one, or precisely $M_{i'j}$ and $M_{ij'}$ are.  Clearly, for any such configuration, by switching the two zero entries to one and vice-versa preserves all row and column sums; Ryser showed that \emph{any} pair of matrices with equal row and column sums can be obtained from one another by a sequence of such rectangular, $0$-$1$ swaps.\medskip
\begin{figure}
\[
M = \begin{bmatrix}
1 & 0 & 1\\
{\color{red} 1} & 1 & {\color{red} 0}\\
{\color{red} 0} & 1 & {\color{red} 1}
\end{bmatrix} \leftrightarrow
\begin{bmatrix}
{\color{red} 1} & {\color{red} 0} & 1\\
{\color{red} 0} & {\color{red} 1} & 1\\
1 & 1 & 0
\end{bmatrix}\leftrightarrow
\begin{bmatrix}
0 & 1 & 1\\
1 & {\color{red} 0} & {\color{red}  1}\\
1 & {\color{red}  1} & {\color{red} 0}
\end{bmatrix} = N
\]
\caption{Two $3\times 3$ matrices, $M$ and $N$, with equal row and column sums. Ryser proved that any such pair of matrices can be obtained from one another via a sequence of `rectangular swaps.'}
\end{figure}

Conversely, any preference $\succ$ may be represented as a $0$-$1$ vector, whose components are indexed by pairs $(x,A)$ with $x \in A \subseteq X$, via $\mathbbm{1}_{\{x \; \succ \,A \setminus x\}}$. Thus sets of preferences $\mathcal{M} \subseteq \mathcal{L}$ define a $0$-$1$ matrix whose columns correspond to the vector representations of preferences in $\mathcal{M}$.  Moreover, two equal-sized sets of preferences have behaviorally equivalent uniform distributions if and only if their matrix representations have equal row and column sums, and hence can be obtained from one another by a sequence of rectangular, $0$-$1$ swaps. However, not all $0$-$1$ matrices arise as sets of preferences in this manner. This means that Ryser's original switching operation is too strong for our problem: it generally results in matrices which do not represent any collection of preferences.\medskip

In contrast, our operation of swapping terminal segments of separable pairs of preferences not only preserves all row and column sums (hence can be realized as a sequence of rectangular $0$-$1$ switches) but also, crucially, preserves the property of being a matrix of preferences. Now, to characterize behavioral equivalence, we are forced to also consider non-uniform distributions, unlike \cite{ryser1957combinatorial}. However, \autoref{ryserspacetheorem} shows that by using such swaps to transfer portions of mass (i.e.\ \emph{weighted} Ryser swaps) between distributions, we are able to obtain a similar characterization, even in our generalized setting.

\section{Support Restrictions}\label{sec:fullSub}

A natural means of restricting the random utility models is to impose ex-ante constraints on the sets of preferences over which randomization occurs.  We say that a model $\mathcal{M} \subseteq \Delta(\mathcal{L})$ is defined by {\bf support restrictions} if there is some subset $\mathcal{S} \subseteq \mathcal{L}$ such that $\mathcal{M} = \{\mu \in \Delta(\mathcal{L}) : \mu \vert_{\mathcal{L} \setminus \mathcal{S}} = 0\}$.  Support restrictions are natural in contexts when the modeller believes there is cause to a priori restrict the sets of preferences which are held in a population, but not the relative frequencies of those preferences.\medskip


Let $\tilde{\Delta}(\mathcal{S}) \subseteq \mathbb{R}^\mathcal{L}$ denote the face of $\Delta(\mathcal{L})$ spanned by the preferences in $\mathcal{S}$. By \autoref{ryserspacetheorem}, a model defined by support restrictions is identified if and only if for all $\mu \in \tilde{\Delta}(\mathcal{S})$, it is the case that $(\mu + \mathcal{R}) \cap \tilde{\Delta}(\mathcal{S}) = \{ \mu\}$. Finally, let $\mathbb{R}^\mathcal{L}_\mathcal{S}$ denote the linear subspace of signed measures that are zero outside $\mathcal{S}$.  Then the following theorem characterizes identifying support restrictions.\medskip

\begin{theorem}\label{thm:FullSubmodelThm}
    Let $\mathcal{M}$ be defined by the support restriction $\mathcal{S} \subseteq \mathcal{L}$. Then the following are equivalent.
    \begin{enumerate}
        \item $\mathcal{M}$ is identified.
        \item For all $\mu \in \tilde{\Delta}(\mathcal{S})$, 
        \[
        (\mu + \mathcal{R}) \cap \tilde{\Delta}(\mathcal{S}) = \{ \mu\}.
        \]
        \item For every finite sequence of \emph{unweighted} Ryser swaps $\{R_i\}_{i=1}^K$ (allowing for repetition):
        \[
            \sum_{i=1}^K R_i \in \mathbb{R}^\mathcal{L}_\mathcal{S}  \iff \sum_{i=1}^K R_i = \mathbf{0}.
        \]
    \end{enumerate}
\end{theorem}

The equivalence between (1) and (2) is simply a restatement of \autoref{ryserspacetheorem}.  In contrast, the equivalence of (3) is only valid for models defined through support restrictions.  Condition (3) says that $\mathcal{M}$ is identified if and only if every (non-zero) vector in $\mathcal{R}$ places weight on some preference outside $\mathcal{S}$. It shows that, in the case of support restrictions, testing for identification does not depend on the weights, but rather only on the specifics of the family of allowable preferences.\medskip

\autoref{thm:FullSubmodelThm} also provides a characterization of linear independence for choice functions induced by preferences.  Any preference $\succ$ defines a random choice rule $\rho_\succ(x,A) = \mathbbm{1}_{\{x \; \succ \,A \setminus x\}}$. The model $\mathcal{M}$ defined by the support restriction $\mathcal{S} \subseteq \mathcal{L}$ is identified if and only if the choice functions $\{\rho_\succ\}_{\succ \in \mathcal{S}}$ are linearly independent. Hence \autoref{thm:FullSubmodelThm} also subsumes the partial characterization of independent choice functions in \cite{chambers2024limits}.

\subsection{Extreme Points of Partially Identified Sets}

Suppose that $\mathcal{M}$ is defined by the support restriction $\mathcal{S} \subseteq \mathcal{L}$.  In light of \autoref{thm:FullSubmodelThm}, in general $\mathcal{M}$ will not be identified. This means that if $\rho$ is rationalizable by some $\mu \in \mathcal{M}$, it is likewise rationalized by any distribution in the polytope $\mathcal{M} \cap \big(\mu + \mathcal{R}\big)$.  The following theorem characterizes the extreme points of this set.

\begin{theorem}\label{thm:extremepoints}
    Let $\mathcal{M}$ be defined via the support restriction $\mathcal{S} \subseteq \mathcal{L}$. For $\mu \in \mathcal{M}$, the following are equivalent.
    \begin{enumerate}
        \item $\mu$ is an extreme point of $\mathcal{M} \cap \big(\mu + \mathcal{R}\big)$.
        \item The choice functions $\{\rho_\succ :  \mu(\succ) > 0\}$ are linearly independent.
        \item For every finite sequence of \emph{unweighted} Ryser swaps $\{R_i\}_{i=1}^K$ (allowing for repetition):
        \[
            \sum_{i=1}^K R_i \in \mathbb{R}^\mathcal{L}_{\textrm{supp}(\mu)}  \iff \sum_{i=1}^K R_i = \mathbf{0}.
        \]
    \end{enumerate}
\end{theorem}

The equivalence between the first two conditions in \autoref{thm:extremepoints} follows from a theorem of \cite{winkler1988extreme}. In contrast, the equivalence between condition (3) and the first two is novel and a consequence of \autoref{thm:FullSubmodelThm}: a distribution is extremal in the set of behaviorally identical measures in $\mathcal{M}$ if and only if no non-zero vector in the Ryser subspace $\mathcal{M}$ places any weight on any preference not in the support of $\mu$. Notably, this does not depend on $\mathcal{S}$ at all, beyond the fact that $\mu \in \mathcal{M}$, and hence $\textrm{supp}(\mu) \subseteq \mathcal{S}$.

\subsection{Relation to Separable Random Choice}

Consider a pair of agents, each of whom chooses over the finite set of alternatives $X$.  We say that a joint random choice rule for these two agents is {\bf stochastically separable} if there exists some joint distribution $\mu \in \Delta\big(\mathcal{L} \times \mathcal{L}\big)$ such that:
\[
    \rho(x,y; \,A,B) = \mu \big\{ (\succ, \succ') : \textrm{$x$ is $\succ$-maximal in $A$ and $y$ is $\succ'$-maximal in $B$}\big\}.
\]
In other words, a joint stochastic choice function $\rho$ is separable if correlation in choice only arises through correlation in the preferences of the agents.\medskip

In general, testing for stochastic separability is technically challenging (see, e.g.\ \cite{li2021axiomatization,kashaev2024entangled, chambers2024correlated}). However, \cite{kashaev2024entangled} show that when $\mu \in \Delta(\mathcal{S} \times \mathcal{S}\big)$, where the choice functions associated with the preferences in $\mathcal{S}$ are linearly independent, the empirical content of separability becomes straightforward to characterize. To date, no characterization of these sets $\mathcal{S}$ has been known.  However, our \autoref{thm:FullSubmodelThm} provides a complete answer to this question, and hence characterizes exactly those support restrictions for which testing stochastic separability becomes easy.\footnote{For a full discussion of the difficulties of testing stochastic separability, we point the reader to \cite{kashaev2024entangled}.} 

\section{Parametric Random Utility Models}\label{sec:Parm}

In practical applications, random utility models are often specified in parametric form.  If $\Theta \subseteq \mathbb{R}^n$ denotes a set of parameters, we refer to a mapping $F: \Theta \to \Delta(\mathcal{L})$ as a {\bf parametric random utility model}.  By \autoref{ryserspacetheorem}, a parametric random utility model fails to be identified if and only if:
\begin{equation}\label{paramunidentified}
    F(\theta') \in F(\theta) + \mathcal{R},
\end{equation}
for distinct $\theta,\theta' \in \Theta$, i.e. if $F(\theta)$ and $F(\theta')$ belong to some common translate of the Ryser subspace. In principle, this is a complete characterization of identification in the parametric context as well. However, in practice, it is often unclear how to verify this simply by inspection of $F$ itself. In this section, we will investigate conditions under which we can obtain \emph{local} tests of identification.  

\subsection{Primitives}

We will restrict ourselves to \emph{smooth} parametric random utility models. For all definitions relating to smooth maps and manifolds, the reader is referred to, e.g.\ \cite{lee2012smooth}. Throughout this section, we will assume the following mild, regularity conditions on primitives:
\begin{itemize}
    \item[(A.1)] {\bf Regularity of Parameterization}: The map $F: \Theta \to \Delta(\mathcal{L})$ is an injective, smooth map from some open, convex set $\Theta \subseteq \mathbb{R}^k$ into $\Delta(\mathcal{L})$.
\end{itemize}
Let $\pi: \mathbb{R}^\mathcal{L} \to \mathcal{R}^\perp$ denote the (orthogonal) projection of signed measures onto the orthogonal complement of the Ryser subspace, $\mathcal{R}^\perp$.\footnote{If $A$ is any matrix with linearly independent columns which form a basis for $\mathcal{R}$, then:
\[
    \pi(\mu)=  \big(I - A[A^\intercal A]^{-1}A^\intercal\big) \mu,
\]
where $I$ denotes the identity matrix.}  Thus $\pi$ may be regarded as mapping each distribution $\mu$ to its equivalence class of behaviorally indistinguishable measures.  We define the set $\mathcal{M}_F$ to be the projection $\pi \, \circ \, F(\Theta)$, i.e.\ the set of behavioral equivalence classes of distributions in the range of $F$. We assume:
\begin{itemize}
    \item[(A.2)] {\bf Regularity of Projection}: $\mathcal{M}_F$ is a smooth, $k$-dimensional manifold.
\end{itemize}

\noindent By \eqref{paramunidentified}, the parametric random utility model $F$ is unidentified if and only if the map $\bar{F} = \pi \vert_{F(\Theta)} \circ F$ fails to be injective; our focus will be on obtaining conditions which ensure this is not the case.
\begin{itemize}
    \item[(A.3)] {\bf Closedness}: The map $\bar{F} : \Theta \to \mathcal{M}_F$ is closed, i.e. $\bar{F}(C)$ is relatively closed in $\mathcal{M}_F$ for every closed $C \subseteq \Theta$.\footnote{One sufficient condition for this is that $\bar{F}$ be \emph{proper}, i.e. $\bar{F}^{-1}(K)$ is compact in $\Theta$ for every compact $K \subseteq \mathcal{M}_F$.}
\end{itemize}

\noindent We will refer to any mapping $F: \Theta \to \Delta(\mathcal{L})$ satisfying (A.1) - (A.3) as a {\bf smooth} parametric random utility model.

\subsection{Parametric Identification}

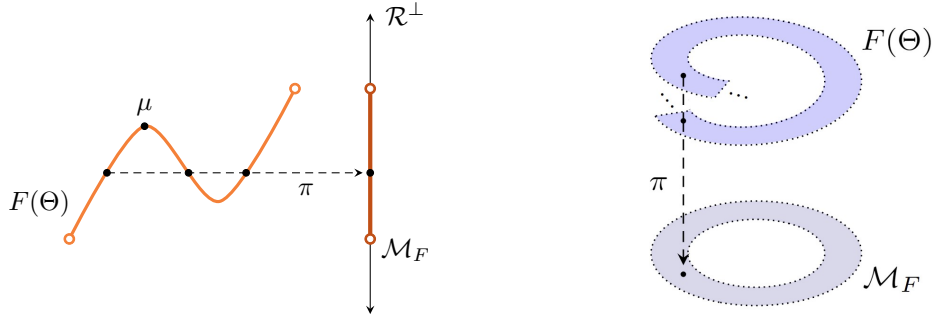
\begin{figure}
    \centering
    \begin{subfigure}[t]{.45\textwidth}
    \centering
        \begin{tikzpicture}

    \draw [Orange, very thick] plot [smooth] coordinates {(0,-.5) (1,1) (2,0) (3,1.5)};

    \filldraw (.5,.38) circle (1.25pt);
    \filldraw (1.585,.38) circle (1.25pt);
    \filldraw (2.35,.38) circle (1.25pt);

    \filldraw (1,1) circle (1.25pt);
    \node[above] at (1,1) {\footnotesize $\mu$};

    \draw[-stealth, densely dashed] (.5, .38) -- (3.9, .38);

    \node[below] at (3.15,.38) {\footnotesize $\pi$};

    \filldraw [Orange] (0,-.5) circle (2pt);
    \filldraw [white] (0,-.5) circle (1pt);

    \filldraw [Orange] (3,1.5) circle (2pt);
    \filldraw [white] (3,1.5) circle (1pt);

    \draw[stealth-stealth] (4,2.5) -- (4,-1.5);

    \draw[Bittersweet, ultra thick] (4,-.5) -- (4, 1.5);
    \filldraw [Bittersweet] (4,-.5) circle (2pt);
    \filldraw [white] (4,-.5) circle (1pt);

    \filldraw [Bittersweet] (4,1.5) circle (2pt);
    \filldraw [white] (4,1.5) circle (1pt);

    \node[right] at (4,-.65) {\footnotesize $\mathcal{M}_F$};

    \node[above] at (-.4,-.35) {\footnotesize $F(\Theta)$};

    \filldraw (4,.38) circle (1.25pt);

    \node[right] (a) at (4,2.5) {\footnotesize $\, \mathcal{R}^\perp$};
    \end{tikzpicture}
    \caption{When $\mathcal{M}_F$ is simply connected, any failure of identification globally implies a local failure, e.g.\ at $\mu$.}
    \end{subfigure}
    \begin{subfigure}[t]{.45\textwidth}
    \centering
    \includegraphics[width=0.55\linewidth]{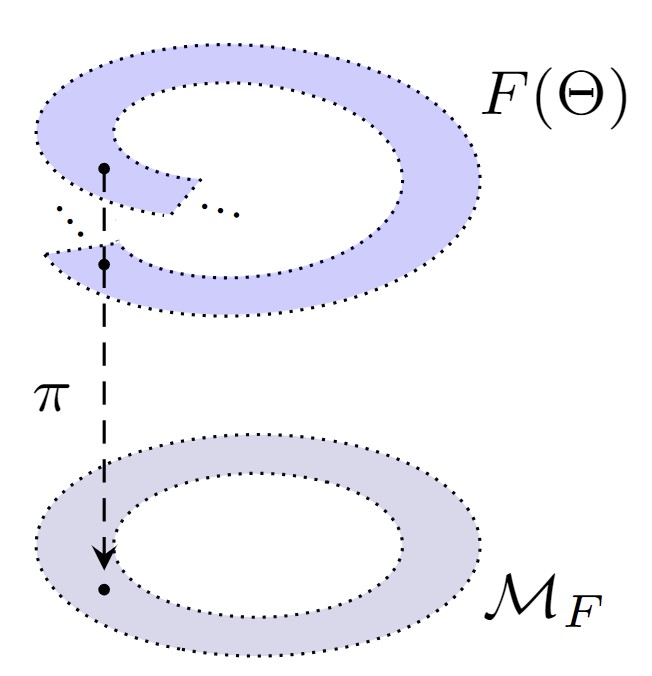}
\caption{When $\mathcal{M}_F$ is not simply connected, it is possible for $\bar{F}$ to be everywhere locally identified, yet still fail to be globally so.}
\end{subfigure}
\caption{An illustration of the role the topology of $\mathcal{M}_F$ plays in guaranteeing the existence of \emph{local} necessary and sufficient conditions for identification.}
\label{simplyconnectedfig}
\end{figure}

We say that a smooth parametric random utility model is {\bf parametrically identified} if the mapping $\bar{F}$ is a diffeomorphism. This simply requires that the map from parameters to behavioral equivalence classes of distributions is a smooth bijection, with a smooth inverse.  Our objective is to obtain conditions under which parametric identification is equivalent to identification holding locally about each parameter.  This is not true for general smooth parametric random utility models, as illustrated by \autoref{simplyconnectedfig}.\medskip

The crucial ingredient in obtaining a local test for identifiability is the topology of $\mathcal{M}_F$.  We say a space is {\bf simply connected} if any closed curve in it can be contracted to a point, without leaving the space.\footnote{Formally, a space $X$ is simply connected if, for every continuous pair $p,q : [0,1] \to X$ of paths such that $p(0) = q(0)$ and $p(1) = q(1)$, there exists a homotopy $H: [0,1]^2 \to X$ such that $H(x,0) = p(x)$ and $H(x,1) = q(x)$ for all $x \in [0,1]$.}  Informally, $\mathcal{M}_F$ is simply connected if and only if it has no `holes' in it. Moreover, since $\Theta$ is convex and hence simply connected,   $\mathcal{M}_F$ must necessary be too if $\bar{F}$ is parametrically identified. However, as our next result shows, whenever $\mathcal{M}_F$ is simply connected, a smooth parametric random utility model $F$ is locally identified if and only if it is so globally. \medskip

\begin{theorem}\label{thm:ParametricRUM}
    Let $F$ be a smooth, parametric random utility model. Suppose that $\mathcal{M}_F$ is simply connected. Then the following are equivalent:
    \begin{itemize}
        \item[(i)]  $F$ is parametrically identified; and
        \item[(ii)] The ($k\times k$) Jacobian matrix $d \bar{F}$ has full rank at every $\theta \in \Theta$.
    \end{itemize}
    
    \end{theorem}

\noindent Despite its seemingly abstract nature, the simply-connectedness of $\mathcal{M}_F$ is critical. As illustrated in \autoref{simplyconnectedfig}, without controlling for the structure of $\mathcal{M}_F$, \autoref{thm:ParametricRUM} is false.\footnote{Concretely, consider the map $F: \mathbb{R} \to S^1$ via $t \mapsto \big(\cos(t), \sin(t)\big)$.  This map has everywhere non-zero differential, but nonetheless is not globally injective. This is possible precisely because the unit circle $S^1 $ fails to be simply connected.}\medskip

In practice, many natural classes of parametric random utility models have a simply connected $\mathcal{M}_F$, allowing us to test for identification globally based purely on local calculations. For models with known or tractable choice probabilities, the following lemma provides a convenient tool for establishing the structure of $\mathcal{M}_F$.

\begin{lemma*}[Structure Lemma]\hypertarget{structurelemma}{}
    For any smooth parametric random utility model  (i) $\Phi \circ F(\Theta)$, the set of random choice rules induced by $F$, and (ii) $\mathcal{M}_F$, are homeomorphic. In particular, one is simply connected if and only if the other is.
\end{lemma*}

\noindent Perhaps the simplest possible examples of where parametric identification can be reduced to a purely local question are are the i.i.d.\ discrete choice models. 

\begin{example}\label{iiddc}
    Suppose $\{\varepsilon_x\}_{x \in X}$ is a collection of i.i.d. random variables with everywhere positive density.  Let:
\[
\Theta = \{\theta \in \mathbb{R}^X : \theta_{\bar{x}} = 0\}
\]
for some fixed $\bar{x}$, and define a parametric random utility model via:
\[
    \mu\{x_1x_2 \cdots x_{\vert X \vert}\} = \mathbb{P}\big(\theta_{x_1} + \varepsilon_{x_i} > \cdots > \theta_{x_{\vert X \vert}} + \varepsilon_{x_{\vert X \vert}}\big).
\]
For example, when the $\varepsilon$'s are distributed extreme value, this specification yields the logit model, and when the $\varepsilon$ are Gaussian, the probit. Any parametric random utility model of this form is fully characterized by its choice probabilities on binary sets; in fact there is a one-to-one correspondence between any tuple of probabilities $\big\{\rho\big(x, \{\bar{x}, x\}\big)\}_{x \in X} \in (0,1)^X$ and distributions in $F(\Theta).$  As a consequence, the \hyperlink{structurelemma}{Structure Lemma} immediately implies that $\mathcal{M}_F$ is simply connected, and hence questions of identification are necessarily of a purely local nature. \hfill $\blacksquare$
\end{example}

While \autoref{iiddc} provides a number of cases in which testing identification is local in nature, it is straightforward to show these models are identified directly, without the need to invoke the machinery of \autoref{thm:ParametricRUM}. An equally broad class of models in which \autoref{thm:ParametricRUM} provides a purely local test of identification are the so-called mixture models. Formally, a parametric random utility model is said to be a {\bf mixture model} if the set of random choice rules it induces is \emph{convex}. Mixture models are natural candidates for describing population-level heterogeneity. Indeed, \cite{strzalecki2024stochastic} writes:
\begin{quote}
    \emph{``[W]e can venture to say that a class [of random choice rules] is not a good model of a population if it is not closed under mixtures."}\footnote{See \cite{strzalecki2024stochastic}, p. 57.}
\end{quote}

\noindent As every convex set is trivially simply connected, as a consequence of the \hypertarget{structurelemma}{Structure Lemma} we obtain that for any parametric mixture model, global identification is testable purely locally.

\begin{corollary}\label{mixturemodelprop}
    Suppose $F$ defines a smooth parametric mixture model. Then $F$ is identified if and only if $d\bar{F}_\theta$ has full rank at every $\theta \in \Theta$.
\end{corollary}

\subsection{The Geometry of Parametric Identification}

Suppose now, that $F$ is itself either an open or closed map. Under either of these conditions, it follows that $F(\Theta)$ itself is a smooth manifold, embedded in $\Delta(\mathcal{L})$.\footnote{E.g.\ \cite{lee2012smooth} Proposition 4.22.}  The tangent space to this manifold, at any point, is given by the column space of the matrix $dF_\theta$.  Now, by the chain rule:
\[
    d\bar{F} = d\big(\pi \vert_{F(\Theta)}\big) \circ dF.
\]
The map $\pi \vert_{F(\Theta)}$ is the restriction of a linear map, hence its differential is simply the map $\pi$ itself. Thus $\bar{F}$ fails to be locally injective, if and only if, at some $\theta$, the tangent space of $F(\Theta)$ at $\mu = F(\theta)$ has non-trivial intersection with the kernel of $\pi$, i.e. $\mathcal{R}$.  In other words, when $\mathcal{M}_F$ is simply connected, $F$ is identified if and only if no translate of $\mathcal{R}$ is tangent to $F(\Theta)$, the set of distributions traced out by the parameterization function $F$. This condition is violated, for example, at the measure $\mu$ in part (a) of \autoref{simplyconnectedfig}. \medskip

\begin{corollary}
    Suppose $F$ is a smooth parametric random utility model, that (i) $F$ is either an open or closed map, and (ii) $\mathcal{M}_F$ is simply connected. Then the following are equivalent:
    \begin{itemize}
        \item[(i)] $F$ is parametrically identified.
        \item[(ii)] The $(k \times k)$ matrix $d\bar{F}$ is everywhere of full rank.
        \item[(iii)] No translate of $\mathcal{R}$ is anywhere tangent to the manifold $F(\Theta)$.
    \end{itemize}
\end{corollary}

\section{Conclusion}\label{sec:conc}

This paper considers the problem of characterizing those ex-ante restrictions on the random utility model which yield identification. Our key observation is that a simple collection of mass-swapping operations, what we term \emph{Ryser swaps}, generate the full class of transformations which preserve all choice probabilities. Given an arbitrary set $\mathcal{M} \subseteq \Delta(\mathcal{L})$, $\mathcal{M}$ is identified if and only every translate of the Ryser subspace intersects $\mathcal{M}$ in at most one point.  When $\mathcal{M}$ corresponds to the set of distributions supported on some subset of preferences $\mathcal{S} \subseteq \mathcal{L}$, we find that a restricted random utility model is identified if and only if every finite sequence of Ryser swaps places some mass outside the support restriction. We  obtain a related characterization of the extremal rationalizing random utility models.\medskip

When models are parametric, we use homotopy-theoretic techniques to reduce the complex, global problem of testing identification for general models to a straightforward, local one: testing the invertibility of a certain Jacobian matrix at each parameter value.  We use this, e.g., to show that the identification of any mixture model is wholly determined by local information.\footnote{While we have not done so in this paper, it is straightforward to extend our techniques to random choice models outside the random utility framework.}\medskip


\bibliographystyle{ecta}
\bibliography{linind}

\appendix

\section{Preliminary Constructions}
In this appendix we introduce some mathematical preliminaries needed for the proofs of our results. To begin, we define the \textbf{M\"{o}bius inverse} of a random choice rule. This construct is also known as the Block-Marschak polynomials of \citet{block1959random}. The M\"{o}bius inverse of a random choice rule $\rho$ is given by a function $q: X \times 2^X\setminus \{\emptyset\} \rightarrow \mathbb{R}$ and is defined as follows.
\begin{equation}
    \begin{split}
        \rho(x,A) & = \sum_{A \subseteq B} q(x,B) \\
        q(x,A) & = \sum_{A \subseteq B} (-1)^{|B \setminus A|}\rho(x,B)
    \end{split}
\end{equation}
We will use the notation $q_\mu$ to denote the M\"{o}bius inverse of $\Phi(\mu)$. In our following proofs, we need to use a result due to \citet{falmagne1978representation} which tell us about the relationship between $q(x,A)$ and the random utility model.

\begin{theorem}[\citet{falmagne1978representation}]\label{thm:falident}
    A distribution over preferences $\mu$ rationalizes a random choice rule $\rho$ if and only if $\mu[U(x,X \setminus A)]=q(x,A)$ for all $x \in A \subseteq X$.
\end{theorem}
\autoref{thm:falident} is just a restatement of Falmagne's theorem from \autoref{sec:geometry} in its original form.

\section{Preliminary Results}
In this section, we provide some preliminary results which are necessary for the proofs of our main results.

\begin{lemma}\label{RyserObs}
    Suppose $\mu$ is a signed probability measures of $\mathcal{L}$ and suppose that $R$ is a weighted Ryser swap, then $\mu$ and $\mu + R$ are observationally equivalent.
\end{lemma}

\begin{proof}
    $\mu$ and $\mu + R$ only differ in their weights on $\{\succ_i,\succ_j,\succ_k,\succ_l\}$. Choice probabilities are linear functions of $\mu(\succ)$. As such, it is sufficient to look at the vector induced by $\mu-(\mu+R)=R$. Note that if these two distributions are observationally equivalent, then $\sum_{\succ \in \mathcal{L}}(\mu(\succ)-(\mu(\succ)+R(\succ)))\mathbf{1}\{x \succ A \setminus \{x\}\}=\sum_{\succ \in \mathcal{L}}-R(\succ)\mathbf{1}\{x \succ A \setminus \{x\}\}$ should be equal to zero for all $x \in A \subseteq X$. Note that $-R$ defines a Ryser swap as $R$ is a ryser swap. By Lemma 4 of \citet{chambers2024correlated} (which is an extension of \autoref{thm:falident} to arbitrary measures), we know that two distributions are behaviorally equivalent if and only if $\mu[U(x,B)]=\nu[U(x,B)]$ for all $x \in B \subseteq X$. By the definition of conjugate squares, we know that the four preferences in the conjugate square, $\{\succ_1,\succ_2,\succ_3,\succ_4\}$, satisfy $|U(x,X \setminus A)\cap \{\succ_1,\succ_2\}|=|U(x,X \setminus A)\cap \{\succ_3,\succ_4\}|$. Further, since each preference in $\{\succ_1,\succ_2,\succ_3,\succ_4\}$ gets equal weight in a Ryser swap, we are done.
\end{proof}

    \begin{lemma}[Zipper Lemma 1]\label{localswap}
        Suppose that $\mu$ and $\nu$ are observationally equivalent. Fix $n \in \{1,\dots,|X|-1\}$. Further suppose that there exist $\succ$ in the support of $\mu$ and $\{\succ_i\}_{i=1}^k$ each in the support of $\nu$ such that $\mu(\succ)\leq \sum_{i=1}^k \nu(\succ_i)$ and $s_{n}^\uparrow(\succ)=s_{n}^\uparrow(\succ_i)$ for each $i$. Then there exists a finite sequence of weighted Ryser swaps applied to $\nu$ resulting in $\nu'$ such that there exists a set of preferences in the support of $\nu'$ given by $\{\succ_j\}_{j=1}^l$ satisfying $s_{n+1}^\uparrow(\succ)= s_{n+1}^\uparrow(\succ'')$ and $\mu(\succ) \leq \sum_{j=1}^l \nu'(\succ_j)$.
    \end{lemma}

    \begin{proof}
        Let $X\setminus A$ be the set equal to the first $n$ elements of $\succ$ and let $x \in A$ be the $n+1$ ranked alternative of $\succ$. By observational equivalence of $\mu$ and $\nu$ and \autoref{thm:falident}, we have that $q_{\nu}(x,A)=q_{\mu}(x,A) \geq \mu(\succ)$. Consider the set of preferences given by $U=U(x,X\setminus A) \cap supp(\nu)$ which is non-empty by the last sentence. Enumerate $U$ via $j \in \{1,\dots,m\}$. For every preference $\succ_j\in U$ and for each $\succ_i$ from the statement of the lemma, note that either $\succ_i \in U$, $\succ_j$ is in the set of preferences from the statement of the lemma, or  $(\succ_i,\succ_j)$ form a conjugate square with $(\succ',\succ'')$ for $\succ'$ and $\succ''$ formed as follows.
        \begin{itemize}
            \item $\succ' = s_n^\uparrow(\succ_i) \cdot s_n^\downarrow(\succ_j)$
            \item $\succ'' = s_n^\uparrow(\succ_j) \cdot s_n^\downarrow(\succ_i)$
        \end{itemize}
        Note that for all such $\succ'$ we have that $s_{n+1}^\uparrow(\succ')=s_{n+1}^\uparrow(\succ)$. This follows as $s_n^\uparrow(\succ)=s_n^\uparrow(\succ_i)$ and because $\succ$ and $\succ_j$ both have $x$ as their $n+1$ ranked alternative. Let $\{(\succ_i,r_i)\}_{i=1}^k$ denote a sequence of pairs where $r_i=\sum_{o=1}^i \nu(\succ_o)$. Similarly, let $\{(\succ_j,r_j)\}_{j=1}^m$ denote a sequence of pairs where $r_j=\sum_{o=1}^j \nu(\succ_o)$. Now let $s^f=(s_f,i_f,j_f)$ such that the following conditions hold.
        \begin{itemize}
            \item If $r_{i_f} > r_{j_f}$, then either $i_{f-1}=i_f$ or ($i_{f-1}=i_f-1$ and $r_{i_{f-1}} \leq r_{j_f}$). Further, $s_f=r_{j_f}$.
            \item If $r_{i_f} \leq r_{j_f}$, then either $j_{f-1}=j_f$ or ($j_{f-1}=j_f-1$ and $r_{j_{f-1}} \leq r_{i_f}$). Further, $s_f=r_{i_f}$.
        \end{itemize}
        Now we translate this setup into words. Our enumeration of $\{\succ_i\}$ and $\{\succ_j\}$ orders these two sets of preferences. Then $r_i$ keeps track of the total probability weight $\nu$ puts on all preferences between $\succ_1$ and $\succ_i$ with a similar statement for $r_j$. Finally, $(s_f,i_f,j_f)$ is a way of comparing $r_i$ and $r_j$. Specifically, $s_f$ is the next smallest value after $s_{f-1}$ among all $r_i$ and $r_j$. If $i_f - i_{f-1}=1$ then this next smallest value is an $r_i$ and if $j_f-j_{f-1}=1$ then this next smallest value is an $r_j$. We now use this sequence of $s_f$ to perform a sequence of weighted Ryser swaps.
        \begin{enumerate}
            \item Initialize at $f=1$, let $s_0=0$, and let $\nu_1=\nu$.
            \item If $\succ_{i_f} \not \in U$ and $\succ_{j_f}$ is not in the set of preferences from the statement of the lemma, add weighted Ryser swap $R$ to $\nu_f$ at weight $s_f-s_{f-1}$ where $R$ is over the conjugate pair $(\succ_{i_f},\succ_{j_f})$ and $(\succ',\succ'')$ where $(\succ',\succ'')$ is the pair of preferences described earlier in the proof.
            \item This addition outputs $\nu_{f+1}$. If we have that either, for all $g,g' \geq f+1$,  $i_g=i_{g'}$ or, for all $g,g' \geq f+1$, $j_g=j_{g'}$, then terminate the algorithm and output $\nu'=\nu_{f+1}$. If not, set $f=f+1$ and return to step 2.
        \end{enumerate}
        Every step of this algorithm is a weighted Ryser swap and thus maintains behavioral equivalence by \autoref{RyserObs}.\footnote{Here we specifically allow for trivial Ryser swaps. By a trivial Ryser swap, we mean a separable pair that maps to itslef.} Further, adding each of these Ryser swaps results in $s_f-s_{f-1}$ weight going onto a preference $\succ'$ such that $s_{n+1}^\uparrow(\succ')=s_{n+1}^\uparrow(\succ)$. The algorithm terminates with the total weight summed across all summed Ryser swaps being equal to $\min\{\sum_{i=1}^k \nu(\succ_i), q(x,A)\}$. Both $\sum_{i=1}^k \nu(\succ_i)$ and $q(x,A)$ are weakly larger than $\nu(\succ)$. Thus, if we let $\{\succ_p\}_{p=1}^l$ enumerate the aforementioned $\succ'$ preferences, we have that $\sum_{p=1}^l \nu'(\succ_p) \geq \nu(\succ)$, and so we are done.
    \end{proof}

    \begin{lemma}[Zipper Lemma 2]\label{swapsupport}
        Suppose $\mu$ and $\nu$ are observationally equivalent. Fix a $\succ$ in the support of $\mu$. There exists a finite sequence of weighted Ryser swaps $\{R_i\}$ such that $\nu + \sum_i R_i = \nu'$ and $\nu'(\succ)\geq \mu(\succ)$.
    \end{lemma}

    \begin{proof}
        Suppose that $\mu$ and $\nu$ are observationally equivalent. Fix some preference in the support of $\mu$ and call it $\succ$. By \autoref{thm:falident}, we have that $q_{\mu}=q_{\nu}$. Specifically, this means that for $x$ which is ranked highest by $\succ$, we have that $q_{\mu}(x,X)=q_{\nu}(x,X)$. This means that there are a collection of preferences in the support of $\nu$, specifically $U_1=U(x,\{x\}) \cap supp(\nu)$, such $\mu(\succ) \leq \nu(U_1)$. We now apply \autoref{localswap} and see that we get there is a collection of preferences, $U_2$, in the support of our induced distribution, $\nu'$, with each $\succ' \in U_2$ satisfying $s_{2}^\uparrow(\succ)=s_{2}^\uparrow(\succ')$. With repeated but finite applications of \autoref{localswap}, we get the same statement replacing $2$ with $|X|$. However, $s_{|X|}^\uparrow(\succ)=s_{|X|}^\uparrow(\succ')$ implies that $\succ$ and $\succ'$ agree on every ranking and thus $\succ=\succ'$. At each step of this application we have maintained $\mu(\succ) \leq \nu'(U_i)$ and thus we have $\mu(\succ) \leq \nu'(U_{|X|})$. As we just mentioned, this is equivalent to $\mu(\succ)\leq \nu'(\succ)$ where $\nu'$ is the induced distribution after repeated application of \autoref{localswap}.
    \end{proof}

\begin{lemma}\label{rationalLem}
    Let $\mathcal{R}$ denote the subspace of $\mathbb{R}^\mathcal{L}$ spanned by vectors of the form:
    \[
        \mathbbm{1}_{\{\succ, \succ'\}} - \mathbbm{1}_{\{\succ'', \succ'''\}}
    \]
    where $(\succ,\succ')$ and $(\succ'', \succ''')$ form a conjugate square, and by minor abuse of notation let $\Delta(\mathcal{S})$ denote the face of $\Delta(\mathcal{L})\subseteq \mathbb{R}^\mathcal{L}$ spanned by the abstract simplex $\varnothing \subsetneq \mathcal{S} \subseteq \mathcal{L}$.  Suppose there exists $t \in \mathbb{R}^\mathcal{L}$ such that:
    \[
       \dim \bigg[\big(\mathcal{R} + t\big) \cap \Delta(\mathcal{S})\bigg] > 0.
    \]
    Then there exists $t' \in \mathbb{R}^\mathcal{L}$ such that $\big(\mathcal{R} + t'\big) \cap \Delta(\mathcal{S})$ contains a pair of distinct points in $\mathbb{Q}^\mathcal{L}$.
\end{lemma}
\begin{proof}
    Let $\mathcal{V}$ denote the subspace of $\mathbb{R}^\mathcal{L}$ given by:
    \[
        \textrm{span } \big\{e^{\succ} - e^{\succ'}\big\}_{\succ,\succ' \in \mathcal{S}},
    \]
    where $e^i$ denotes the $i$th standard Euclidean  basis vector.  By hypothesis, 
    \[
        K = \textrm{dim } \mathcal{R} \cap \mathcal{V} > 0.
    \]
    From their definitions, both $\mathcal{R}$ and $\mathcal{V}$ admit bases $\{q_{\mathcal{R}}^i\}_{i = 1}^{\textrm{dim}(\mathcal{R})}$ and $\{q_{\mathcal{V}}^j\}_{j = 1}^{\textrm{dim}(\mathcal{V})}$ which belong to $\mathbb{Q}^\mathcal{L}$. Define the $\vert \mathcal{L}\vert  \times \textrm{dim}(\mathcal{R})$ and $\vert \mathcal{L}\vert \times \textrm{dim}(\mathcal{V})$ matrices:
    \[
        Q_\mathcal{R} = \begin{bmatrix} q^1_\mathcal{R} & \cdots & q^{\textrm{dim}(\mathcal{R})}_\mathcal{R} \end{bmatrix}
    \]
    and
    \[
        Q_\mathcal{V} = \begin{bmatrix} q^1_\mathcal{V} & \cdots & q^{\textrm{dim}(\mathcal{V})}_\mathcal{V} \end{bmatrix},
    \]
    respectively, and let:
    \[
        Q = \begin{bmatrix} Q_\mathcal{R} & \aug & -Q_\mathcal{V} \end{bmatrix}
    \]
    Since each of the above matrices consists exclusively of rational elements, their respective column spaces trivially admit a bases in $\mathbb{Q}^\mathcal{L}$. This implies that, by Gaussian elminiation, the annihilators of their column spaces admit a bases in $\mathbb{Q}^\mathcal{L}$; in particular the annihilator of the column space of $Q$ admits a rational basis $\{r^K\}_{k=1}^K$, where each vector $r^k = [r^k_\mathcal{R} \; \vert \; r^k_\mathcal{V}]$.\medskip

    Let $\bar{q}^k = Q_\mathcal{R} r^k_\mathcal{R} \, \big( = Q_\mathcal{V} r_\mathcal{V}^k\big)$.  By construction, $\{\bar{q}^k\}_k \subset \mathbb{Q}^\mathcal{L}$ and, by construction $\{\bar{q}^k\}_k$ form a basis for $\mathcal{R} \cap \mathcal{V}$. Since $\textrm{dim}(\mathcal{R} \cap \mathcal{V}) > 0$, there exists some non-zero $\bar{q}^k \in \mathcal{R} \cap \mathcal{V} \cap \mathbb{Q}^\mathcal{L}$. Then letting:
    \[
        t' = \frac{1}{\vert \mathcal{S}\vert} \mathbbm{1}_{\mathcal{S}},
    \]
    we have $t'$ and $t' + \alpha \bar{q}^k$, for small enough $\alpha \in \mathbb{Q}_{++}$, are distinct rational vectors in $\big(\mathcal{R} + t'\big) \cap \Delta(\mathcal{S})$ as desired.
\end{proof}

\section{Theorem Proofs}

\subsection{Proof of \autoref{ryserspacetheorem}}

We begin by stating and proving an equivalent result.

\begin{lemma}\label{MainLemma}
    Two distributions $\mu$ and $\nu$ are observationally equivalent if and only if there exists a finite sequence of weighted Ryser swaps $\{(R_i,w_i)\}$ such that $\mu + \sum_i w_i R_i = \nu$.
\end{lemma}

\begin{proof}
    Suppose that $\mu + \sum_i w_i R_i = \nu$ where $\{(R_i,w_i)\}$ is a finite sequence of weighted Ryser swaps. By repeated application of \autoref{RyserObs}, we get that $\nu$ is observationally equivalent to $\mu$. We now prove necessity. To do so, we now construct an algorithm to go from $\nu$ to $\mu$ via a sequence of weighted Ryser swaps.
    \begin{enumerate}
        \item Initialize by enumerating the set of preferences in the support of $\mu$ via $i \in \{1, \dots, n\}$ and set $i=1$. Set $\nu_i=\nu$ and set $\mu_i=\mu$. 
        \item $\nu_i$ is observationally equivalent to $\mu_i$. As such we can fix $\succ_i$ in the support of $\nu_i$ and apply \autoref{swapsupport} to $\mu_i$ to get $\mu_i'$ such that $\mu_i'(\succ_i)\geq \nu_i(\succ_i)$.
        \item Set $i=i+1$. Set $\mu_i(\succ)=\mu_{i-1}(\succ)-\nu(\succ)\mathbf{1}\{\succ=\succ_{i-1}\}$. Set $\nu_i(\succ)=\nu_{i-1}(\succ)-\nu(\succ)\mathbf{1}\{\succ=\succ_{i-1}\}$.
        \item If $\nu_i=\mathbf{0}$, terminate the algorithm. If not, return to step 2.
    \end{enumerate}
    Recall that addition of weighted Ryser swaps maintains observation equivalence. Further, at every iteration of the above algorithm, if we subtract out $\nu(\succ)$ from one distribution, we are also subtracting out $\nu(\succ)$ from the other distribution. Thus at every iteration of the algorithm we are maintaining observational equivalence (subject to not summing to one). Since we are subtracting $\nu(\succ_i)$ from $\nu(\succ_i)$ at each iteration of the algorithm and iterating over each $i$, the algorithm terminates when $i$ reaches $n+1$. Again, since observational equivalence was maintained at every iteration of the algorithm, $\mu_{n+1}$ is also equal to $\mathbf{0}$. Note that in this algorithm we only apply \autoref{swapsupport} finitely many times. Further, the proof of \autoref{swapsupport} only applies \autoref{localswap} a finite number of times and \autoref{localswap} only uses a finite number of weighted Ryser swaps. Finally, this means that the finite sequence of weighted Ryser swaps implied by the algorithm take $\nu$ to $\mu$, and thus we are done.
\end{proof}

Now we can state the proof of \autoref{ryserspacetheorem}.

\begin{proof}
    Observe that $\{\sum_i w_i R_i| w_i \in \mathbb{R}\}$ defines $\mathcal{R}$. So $\mu + \mathcal{R}$ is the set $\{\mu+\sum_i w_i R_i| w_i \in \mathbb{R}\}$. This is exactly the set of measure observationally equivalent to $\mu$ by \autoref{MainLemma}. By definition, $\mathcal{M}$ is identified if and only if there are no distributions $\nu \in \mathcal{M}$ observationally equivalent to $\mu$ for all $\mu$. This is equivalent to $\big(\mu + \mathcal{R}\big) \cap \mathcal{M} = \{\mu\}$ for all $\mu \in \mathcal{M}$, and so we are done.
\end{proof}

Observe that \autoref{marginalequiv} and \autoref{identifiedsetcorr} follows immediately from \autoref{MainLemma}.

\subsection{Proofs from \autoref{sec:fullSub}}
We begin with our proof of \autoref{thm:FullSubmodelThm}.

\begin{proof}
    We begin with the equivalence between (1) and (2). Observe that $D_{\Delta(\mathcal{S})}$ takes the form $\{\mu-\nu|\mu,\nu \in \Delta(\mathcal{S})\}$. Consider any $\mathbf{0} \neq d \in D_{\Delta(\mathcal{S})} \cap \mathcal{R}$.
    \begin{equation*}
        \begin{split}
            d & = \mu - \nu \\
            & = \sum_i w_i R_i
        \end{split}
    \end{equation*}
    The first equality is by definition and the second equality is by inclusion in $\mathcal{R}$. Thus $\mu=\nu - \sum_i w_i R_i$ where $w_i$ corresponds to the weights on their respective Ryser swaps. By \autoref{ryserspacetheorem}, such $\mu$ and $\nu$ exist if and only if $\Delta(\mathcal{S})$ is not identified. Thus (1) and (2) are equivalent.

    We now show the equivalence between (1) and (3). By \autoref{rationalLem}, it follows that, for some $\mu \in \Delta(\mathcal{S}$), $(\mu + \mathcal{R})\cap \Delta(\mathcal{S}) \neq \{\mu\}$ if and only if there exists some potentially different $\nu \in \Delta(\mathcal{S}) \cap \mathbb{Q}^\mathcal{L}$, such that $(\nu + \mathcal{R})\cap \Delta(\mathcal{S}) \cap \mathbb{Q}^\mathcal{L} \neq \{\nu\}$. Now observe that if we can find two observationally equivalent measures in $\mathbb{Q}^\mathcal{L}_\mathcal{S}$, we can always renormalize them so that they lie in $\Delta(\mathcal{S})$. Further observe that each non-zero element of $\mathcal{R}$ defines two observationally equivalent measures. The first measure corresponds to the positive components and the second corresponds to the negative components (once they are made positive). This tells us that we are identified if and only if we can find two observationally equivalent measures with rational components whose supports lie in $\mathcal{S}$. Call these two measures $\mu$ and $\nu$. Observe that $\mu-\nu$ is also a rational element of $\mathcal{R}$. Thus $\mu- \nu$ can be written as $\sum_i \frac{c_i}{d_i} R_i$. Multiply through by each $d_i$ and we get $b (\mu-\nu)=\sum_i c_i R_i$ where $b$ and each $c_i$ are integers. Thus $\sum_i c_i R_i$ corresponds to a finite sequence of (unweighted) Ryser swaps where $R_i$ is repeated $c_i$ times. Thus we have identification if and only if such a sequence exists. This shows the equivalence of (1) and (3), and so we are done.
\end{proof}

We now move on to our proof of \autoref{thm:extremepoints}.

\begin{proof}
    The equivalence of (1) and (2) follows as an immediate consequence of the main theorem of \cite{winkler1988extreme}. Further, identification of full submodels is equivalent to linear independence of the choice functions induced by the generating set of preferences $\mathcal{S}$. Thus checking for linear independence of $\{\rho_\succ|\succ \in supp(\mu)\}$ is equivalent to checking whether the full submodel $\Delta(supp(\mu))$ is identified. By \autoref{thm:FullSubmodelThm}, this is equivalent to condition (3) and so we are done.
\end{proof}

\section{Parametric Identification Results}\label{parametricappendix}

\subsection{A Primer on Covering Spaces}

Let $X, Y$ be topological spaces.  A map $f: X \to Y$ is said to be a {\bf covering map} if, for every $y \in Y$, there exists an open neighborhood $V$ such that $f^{-1}(V)$ is a disjoint union of open sets in $X$, each of which is mapped by $f$ homeomorphically onto $V$. The space $X$ is referred to as a {\bf covering space} for $Y$; the disjoint open sets making up $f^{-1}(V)$ are called the {\bf sheets} of $X$ over $V$; when $V$ is connected these are precisely the connected components of $f^{-1}(V)$. The following result provides sufficient conditions on a transformation to be a covering map.

\begin{theorem*}[\citealt{browder1954}, Theorem 7]
    Let $X,Y \subseteq \mathbb{R}^n$ be connected, and let $F: X \to Y$ be a local homeomorphism. Then if $F$ is a closed map, it is also a covering map.
\end{theorem*}

If $F: X \to Y$ is a covering map for $Y$, the topology of $Y$, as captured by its fundamental group (for definitions, see \citealt{hatcher2002algebraic}), imposes algebraic constraints on the number of sheets in $X$ (e.g. \citealt{hatcher2002algebraic} Proposition 1.32).  The following is a consequence of this more general structure.

\begin{theorem*}[\citealt{spanier1989algebraic}, Theorem 2.3.9]
        Let $F:X \to Y$ be a covering map, with $Y$ connected and simply connected. Then $F$ is a homeomorphism.
\end{theorem*}

\subsection{Proof of \autoref{thm:ParametricRUM}}

\begin{proof}
    Clearly (i) $\implies$ (ii) hence we will show the converse. Suppose then that $d \bar{F}$ has full rank at every $\theta \in \Theta$.  Given (A.1) and (A.2), by the inverse function theorem for manifolds (e.g.\ \citealt{lee2012smooth} Theorem 4.5), $\bar{F}$ is locally a diffeomorphism.  Since $\mathcal{M}_F$ is a continuous image of a connected space, it is connected; moreover, by (A.3), $\bar{F}$ is closed hence by Theorem 7 of \cite{browder1954}, $\bar{F}$ is a covering map.  Now, $\mathcal{M}_F$ is simply connected by hypothesis. Thus by Theorem 2.3.9 of \cite{spanier1989algebraic},  $\bar{F}$ is a homeomorphism.  Any homeomorphism that is also a local diffeomorphism is a (global) diffeomorphism (e.g.\ \citealt{lee2012smooth} Proposition 4.33) and hence $F$ is parametrically identified. \end{proof}

\subsection{Proof of the Structure Lemma}

\begin{lemma}\label{homeolemma}
    The set $\mathcal{M}_F$ is homeomorphic to $\Phi \circ F(\Theta)$.
\end{lemma}
\begin{proof}
    Let $Y = F(\Theta) \subset \Delta(\mathcal{L})$, and let $U \subseteq Y$ be open and $\mathcal{R}$-saturated, i.e. 
    \[
        U = Y \cap (U + \mathcal{R}).
    \]
    Since $U$ is relatively open in $Y$, $U = U' \cap Y$ for some open $U' \subseteq \mathbb{R}^\mathcal{L}$; since $U' + \mathcal{R}$ is open we may without loss suppose that $U'$ is saturated, i.e. $U' = U' + \mathcal{R}$.\footnote{$U' + \mathcal{R} = \bigcup_{\sigma \in \mathcal{R}} (U' + \sigma)$ where each $U' + \sigma$ is open, and hence is open as a union of open sets.}  Let $Z$ denote the range of $\Phi$, i.e. $Z = \Phi(\mathbb{R}^\mathcal{L})$. As $\Phi$ is a linear surjection from $\mathbb{R}^\mathcal{L}$ to $Z$, it is open, hence $\Phi(U')$ is open in $Z$. Now, by definition, we have:
    \[
        \Phi(U) = \Phi(Y \cap U') \subseteq \Phi(U') \cap \Phi(Y).
    \]
    Suppose then that $\rho \in \Phi(U') \cap \Phi(Y)$.  Then there exists $\mu_\rho \in U'$ and $\nu_\rho \in Y$ such that $\Phi(\mu_\rho) = \Phi(\nu_\rho) = \rho$; by \autoref{ryserspacetheorem}, $\sigma = \nu_\rho - \mu_\rho \in \mathcal{R}$. But since $\nu_\rho = \mu_\rho + \sigma$ for $\mu_\rho \in U'$ and $\sigma \in \mathcal{R}$, we obtain that $\nu_\rho \in U'$, as $U'$ is assumed saturated.  Thus $\nu_\rho \in Y \cap U' = U$ and hence $\rho \in \Phi(U)$ as desired.  We conclude that in fact:
    \[
        \Phi(U) = \Phi(U') \cap \Phi(Y)
    \]
    and hence $\Phi(U)$ is open in $\Phi(Y)$; since $U \subseteq Y$ was an arbitrary saturated open set, we conclude that $\Phi \vert_Y : Y \to \Phi(Y)$ is a quotient map.\medskip

    Let $\sim$ denote the binary relation on $Y$ defined by :
    \[
        \mu \sim \nu \iff \mu - \nu \in \mathcal{R}.
    \]
    It is straightforward to show $\sim$ is an equivalence relation on $Y$. Moreover, by \autoref{ryserspacetheorem} $\mu \sim \nu \iff \Phi(\mu) = \Phi(\nu)$. By Corollary 22.3 in \cite{munkres2014topology}, $\Phi(Y)$ is homeomorphic to $Y/\sim$, endowed with the quotient topology.\medskip

    Consider now $\pi \vert_Y : Y \to  \pi(Y) \equiv \mathcal{M}_F$. By an analogous argument we obtain that $\pi(Y)$ is homeomorphic to $Y/\sim$ and the claim follows.
\end{proof}

\subsubsection{Proof of \autoref{mixturemodelprop}}

\begin{proof}
    Suppose $F$ is a mixture model. Then $\Phi \circ F(\Theta)$ is convex.  By \autoref{homeolemma}, this space is homeomorphic to $\mathcal{M}_F$. Since (i) convex sets are simply connected, and (ii) simply connectedness is a homeomorphism invariant, the result follows immediately.
\end{proof}

\section{Relation Between the Linear Order Polytope and the Conjugate Product Structure}\label{App:Polytope}

In this appendix we discuss the relationship between the conjugate product structure (see \autoref{fig:ProdAndSimp}) and the graph or one-skeleton of the linear order polytope. Specifically, we describe how one can start with a graph representing the conjugate product structure and then recover the graph of the linear order polytope. We begin with a description of the graph of the linear order polytope. For a full description of the graph of the linear order polytope, see \citet{doignon2023adjacencies}. The nodes of this graph correspond to the vertices of the polytope, in this case the linear orders. Two nodes are connected if the corresponding vertices are adjacent. \citet{doignon2023adjacencies} characterizes which vertices are adjacent. They show that two vertices are adjacent if the corresponding linear orders have a common nontrivial initial segment ($s_k^{\uparrow}$ for $k\geq 2$) and a differing terminal segment $(s_k^\downarrow)$ or if they have a common nontrivial terminal segment ($s_k^\downarrow$ for $k \geq 2$) and a differing initial segment ($s_k^\uparrow)$.

Consider the following bipartite graph. There is one node for each (non-trivial) initial segment $s_k^\uparrow$ and each (non-trivial) terminal segment $s_k^\downarrow$. Label each node by $s_k^\uparrow$ or $s_k^\downarrow$ respectively. There exists a edge connecting $s_k^\uparrow$ and $s_l^\downarrow$ if $k=l$ and $s_k^\uparrow \cdot s_k^\downarrow$ forms a preference over $X$. Label this edge as $(\succ,k)$ where $\succ = s_k^\uparrow \cdot s_k^\downarrow$. Call a node initial if it is labeled by an initial segment. Terminal nodes are defined analogously. Conjugate squares exactly correspond to four cycles in this graph. We call the graph constructed this way the \textbf{conjugate graph}.

\begin{figure}
    \centering
\begin{tikzpicture}[scale=.5, transform shape]
    \Large
    \tikzstyle{every node} = [rectangle]
    
        \node (a) at (0,0) {$ab$};
        \node (b) at (0,6) {$ba$};
        
        \node (c) at (6,0) {$cd$};
        \node (d) at (6,6) {$dc$};

        \draw [-] (a) -- (c) node[midway, below, sloped] {$(abcd,2)$};
        \draw [-] (a) -- (d) node[pos=.25, below, sloped] {$(abdc,2)$};  
        \draw [-] (b) -- (c) node[pos=.25, below, sloped] {$(bacd,2)$};
        \draw [-] (b) -- (d) node[midway, below, sloped] {$(badc,2)$};

    \end{tikzpicture}
    \caption{The conjugate graph restricted to the Fishburn example in \autoref{fig:ProdAndSimp}. Note that the nodes correspond to our marginals and the edges correspond to the joints.}
    \label{fig:productflowdiagram}
\end{figure}
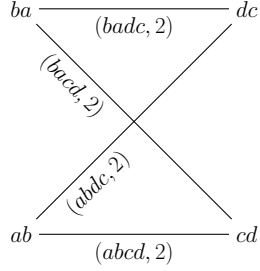

We can think of assigning flows from initial nodes to terminal nodes and representing random utility as a flow feasibility problem. If we let $f$ define a flow, then existence of a flow satisfying the following conditions is equivalent to the existence of a random utility representing the underlying random choice rule.
\begin{equation*}
    \begin{split}
        f \geq 0 &\\
        f(\succ,k) = f(\succ,l) & \forall k,l \text{ }\forall \succ \\
        \sum_{\succ \in U(x,X \setminus A)} f(\succ) = q(x,A)  & \forall x\in A \subseteq X
    \end{split}
\end{equation*}
The set of feasible flows defines our set of observationally equivalent distributions. Now we consider the line graph of this graph. The line graph of a graph is formed by creating one node for each edge on the original graph. Two nodes on the line graph are connected if the two corresponding edges on the original graph share a common node. We then have nodes indexed by $(\succ,k)$. An edge connects $(\succ,k)$ and $(\succ',l)$ if $k=l$ and one of the following holds.
\begin{enumerate}
    \item $s_k^\uparrow(\succ) = s_k^\uparrow(\succ')$ in which case the edge is labeled by $s_k^\uparrow$
    \item $s_k^\downarrow(\succ) = s_k^\downarrow(\succ')$ in which case the edge is labeled by $s_k^\downarrow$
\end{enumerate}
On the line graph of the conjugate graph, a Ryser swap corresponds to four nodes and four edges. Let $n_i$ denote a node and $e_i$ denote an edge.
\begin{enumerate}
    \item $(n_1,n_2,n_3,n_4)$
    \item $(e_1,e_2,e_3,e_4)$ such that $e_1$ connects $(n_1,n_2)$, $e_2$ connects $(n_2,n_3)$, $e_3$ connects $(n_3,n_4)$, and $e_4$ connects $(n_4,n_1)$.
    \item $e_1$ and $e_3$ are initial segments
    \item $e_2$ and $e_4$ are terminal segments
\end{enumerate}
Now there are many nodes labeled with $(\succ,k)$ and $(\succ,l)$. We can construct a multigraph by taking our line graph and combining every node of the form $(\succ,\dots)$. That is to say, every node which is labeled by the same preference is combined. All the edges from our line graph stay the same. Specifically, if an edge $s_k$ connected $(\succ,k)$ and $(\succ',k)$ in the line graph, then an edge with the same label connects $\succ$ with $\succ'$ in this multigraph. We call the multigraph formed this way the \textbf{conjugate inverse multigraph}. We can also represent Ryser swaps on the conjugate inverse multigraph. Given an edge $e_i$ labeled $s_k$, let $e_i(k)$ correspond to the subscript of the label of $e_i$. A Ryser swap corresponds to four nodes and four edges.
\begin{enumerate}
    \item $(n_1,n_2,n_3,n_4)$
    \item $(e_1,e_2,e_3,e_4)$ such that $e_1$ connects $(n_1,n_2)$, $e_2$ connects $(n_2,n_3)$, $e_3$ connects $(n_3,n_4)$, and $e_4$ connects $(n_4,n_1)$.
    \item $e_1$ and $e_3$ are initial segments
    \item $e_2$ and $e_4$ are terminal segments
    \item $e_i(k)=e_j(k)$ for all $i,j$
\end{enumerate}
We need to add the fifth condition for the conjugate inverse multigraph as this fifth condition makes sure that the four segments corresponding to our four edges induce the conjugate product structure shown in \autoref{fig:ProdAndSimp}. 

\begin{definition}
    The \textbf{condensation} of a multigraph is the graph formed by eliminating multiple edges, that is, removing all but one of the edges with the same endpoints.
\end{definition}

\begin{observation}
    The condensation of the conjugate inverse multigraph is the graph of the linear order polytope.
\end{observation}

In both the conjugate inverse multigraph and the graph of the linear order polytope, we have one node for each preference $\succ$. On the graph of the linear order polytope, there is an edge connecting two nodes if they are adjacent. Two preferences are adjacent if and only if they have a common nontrivial initial segment or they have a common nontrivial terminal segment. The conjugate inverse multigraph has an edge between two nodes $\succ$ and $\succ'$ if there exists some $k$ such that $s_k^\uparrow$ is common to $\succ$ and $\succ'$ or if $s_k^\downarrow$ is common to $\succ$ and $\succ'$. It then follows that the graph of the linear order polytope is the condensation of the conjugate inverse multigraph. Intuitively, the reason why we need to introduce the conjugate product structure instead of directly working with the graph of the linear order polytope is because the graph of the linear order polytope fails to keep track of the length of the common initial/terminal segments of two adjacent preferences. The conjugate product relies on initial/terminal segments being of the same length.

We offer one more visualization of this observation. Suppose we start with the line graph of the conjugate graph and we wish to represent this line graph in three-dimensional space. Specifically, we assign nodes with label $(\succ,k)$ a height equal to $k$. We then assign each node with label $(\succ,\dots)$ the same location on their corresponding plane. In other words, no matter the height $k$, each node/preference is found at the same location on a map. The graph of the linear order polytope is then just a top-down view of this three-dimensional representation of the line graph of the conjugate graph.

\end{document}